\documentclass[structabstract]{aa}
\usepackage{natbib}
\bibpunct{(}{)}{;}{a}{}{,}

\usepackage{graphicx}
\usepackage{lscape}
\usepackage{subfigure}
\usepackage{txfonts}
\usepackage{comment}
\usepackage{color}
\usepackage{ifthen}

\newcommand{\Ha}{\mbox{H$\alpha$}}
\newcommand{\Hb}{\mbox{H$\beta$}}

\specialcomment{nccomments}{\begingroup\sffamily\color{blue}}{\endgroup}

\begin{document}

\title{Integral field observations of the blue compact galaxy Haro14\thanks{Based on observations made with ESO Telescopes at Paranal Observatory under program ID 079.B-0445.}}

\subtitle{Star formation and feedback in dwarf galaxies}

\author{L. M. Cair\'os
    \inst{1}
    \and
    J.N. Gonz\'alez-P\'erez
    \inst{2}
    }
\institute{Institut f{\"u}r Astrophysik, Georg-August-Universit{\"a}t,
Friedrich-Hund-Platz 1, D-37077 G{\"o}ttingen, Germany \\
           \email{luzma@astro.physik.uni-goettingen}
           \and Hamburger Sternwarte,
            Gojenbergsweg 112,
21029 Hamburg, Germany\\
\email{jgonzalezperez@hs.uni-hamburg}}

\date{September 2016}

\abstract 
{Low-luminosity, gas-rich blue compact galaxies (BCG) 
 are ideal
 laboratories
to investigate the triggering and propagation of star formation in galaxies, the effects
of massive stellar feedback within a shallow gravitational potential, and the enrichment of the
interstellar medium.}
{We aim to probe the morphology, stellar content, and kinematics, along with 
the nebular excitation and ionization mechanism, in
the BCG Haro\,14 by means of integral field observations.}
{We observed Haro\,14 at the Very Large Telescope, working with the Visible
Multi-Object Spectrograph. From these data we build maps in continuum and in
the brighter emission lines, produce
line-ratio maps (interstellar extinction, density, and diagnostic-line ratios), and obtain  the
velocity and velocity dispersion fields.  We also generate  the integrated spectrum of the major
\ion{H}{ii} regions and young stellar clusters identified in the maps to determine reliable physical
parameters and oxygen abundances.}
{We find as follows: i) the current star formation in Haro\,14 is spatially extended 
with the major \ion{H}{ii} regions placed
along a linear (chain-like) structure, elongated in the north-south direction,
and in a horseshoe-like curvilinear feature that extends about 760~pc eastward; the continuum emission 
is more concentrated and peaks close to the galaxy center; ii) two different episodes
of star formation are present in the central galaxy regions: the recent
starburst, with ages $\leq$ 6~Myrs and the intermediate-age clusters, with ages
between 10 and 30~Myrs; these stellar components rest on a several Gyr old 
underlying host galaxy;     
iii) the H$\alpha$/H$\beta$ pattern is inhomogeneous,
with excess color values varying from $E(B-V)$=0.04 up to $E(B-V)$=1.09; 
iv) shocks play a significant role in the galaxy; and v)  
the velocity field displays a complicated pattern with regions of material moving toward us in the
east and north galaxy areas.}
{The morphology of Haro\,14, its irregular velocity field, 
and the presence of shocks speak in favor of a scenario 
of triggered star
formation. Ages of the knots, i.e., 6~Myr for the starburst  
and 10-30~Myrs for the central clusters, are consistent with the ongoing
burst being triggered by the collective action of stellar winds and supernovae originated in 
the central clusters.}


\maketitle
%

\section{Introduction}

Dwarf galaxies have become very popular objects in contemporary astronomy, and
deservedly so because these small and elusive systems play an essential role in the
understanding of various fundamental astrophysical problems. Dwarf galaxies 
constitute  an unparalleled link to the early Universe since within the framework of a
cold dark matter Universe, galaxies form by hierarchical merger of smaller objects
\citep{Springel2006}. As building blocks of more massive systems, dwarfs provide key
clues in the processes involved in galaxy assembly and evolution.

Among dwarfs, the low luminosity ($M_{B}\geq$-18) and low metallicity
(1/50~Z$\odot\leq$~Z~$\leq$1/2~Z$\odot$; \citealp{Kunth2000}) blue compact galaxies (BCG) appear
as ideal laboratories to probe the still poor understood process of star formation (SF) at large
scales. They are gas-rich galaxies, experiencing intense ongoing SF. Because they are small and
structurally simple, BCGs can neither sustain density waves nor suffer from disk instabilities
and, therefore, offer a unique chance to investigate the SF in an uncomplicated environment.

The study of the effects of energy feedback from massive stars into the surrounding medium in BCGs is
also crucial. Due to their shallow potentials, it has been suggested that dwarf systems can actually
be losing mass \citep{Dekel1986}. If this were the case, this would have important consequences for the
picture of galaxy formation and evolution, process of  SF, and enrichment of the intergalactic 
medium (IGM).  Furthermore, understanding the feedback process will help to establish
the role of star-forming galaxies in the reioinization of the IGM in the early
Universe \citep{Grimes2007}.

\smallskip

Motivated by the importance of these issues we initiated a project focused on the investigation of
BCGs and, in particular, on their ongoing SF episode and the impact of their massive, young stars
on the surrounding interstellar medium (ISM). To this end we undertook a comprehensive analysis of
a sample of about 40 objects by means of integral field spectroscopy (IFS). Integral field unit
(IFU) instruments are particularly suited to probe small, highly asymmetric and compact systems
such as BCGs because in just one shot they provide spectrophotometric and kinematic information on
a large portion of the galaxy; generally the whole starburst is covered. A detailed description of the project as well as our first
results  can be found in  \cite{Cairos2009-mrk1418,Cairos2009-mrk409,Cairos2010, Cairos2012,
Cairos2015}.

\smallskip

The outcomes of this work posed some puzzling new questions and raised the need for further
analyses; in particular, the results found for several galaxies warrant additional work. While analyses of large datasets are fundamental to attack a new field of research and a prerequisite to devise new ideas and theories, the further confrontation with the
individual characteristics of particular objects is the appropriate way to validate the
conclusions.

\smallskip

In this paper we present a detailed analysis of the galaxy Haro\,14. This is a nearby dwarf
system  (M$_{B}$=-16.92), which appears classified as a BCG after its inclusion in the
pioneering work by  \cite{ThuanMartin1981}.  Broadband optical and near-infrared (NIR) data
have been presented in   
\cite{Marlowe1997,Doublier1999,GildePaz2003,GildePaz2005,HunterElmegreen2006,Doublier2001} and
\cite{Noeske2003}.  Optical integrated spectrophotometry has been carried out by \cite{Moustakas2006}
and \cite{Hunter1999}.  We observed Haro\,14, together with seven other
BCGs, working with the \emph{VIsible Multi-Object Spectrograph} (VIMOS;
\citealp{LeFevre2003}).  First results for the whole sample, including emission and
diagnostic-line maps, interstellar extinction and electron density maps, and velocity and
velocity dispersion fields, were presented in \cite{Cairos2015}.  Our analysis revealed
Haro\,14 as a very intriguing object. Its morphology, with the continuum emission concentrated
in the central galaxy regions and the current SF activity taking place in several knots
distributed across the whole observed field, suggests the presence of different episodes of SF.
We identify a bubble-like curvilinear feature, extending about 760~pc from the continuum
peak, and measure high values of [\ion{S}{ii}]~$\lambda\lambda6717,\,6731$/\Ha\ in the galaxy
outskirts, signature of the presence of shocks. These findings motivated further work on this
object, whose results are presented in this paper.

\begin{table}
\caption{Basic parameters for Haro\,14 and collection of data from the literature.
\label{tab:data}}
\begin{center}
\begin{tabular}{lcc}
\hline\hline
Parameter   & Data & Reference \\ 
\hline
\\
Other names        & NGC 0244, UGCA 10, & \\
 & VV~728, PGC~2675  &      \\
 RA (J2000)        &  00$^h$45$^m$46$\fs$4  &      \\
 DEC (J2000)       & -15$\degr$35$\arcmin$49$\arcsec$     &      \\
 V$_{heliocentric}$        &  941$\pm$0.06 km~s$^{-1}$ & \\
 Distance                 &  13.0$\pm$0.09      Mpc  &     \\
 D$_{25}$      &    1.09$\pm$0.06 arcmin   &    RC3   \\
 A$_{B}$       &       0.075                    &     \\
 Morphology        & SOpec; BCG    & RC3; TM81\\\hline
 m$_{B}$          & 13.65$\pm$0.05$^{a}$              &  GP03 \\
 m$_{R}$          &12.91$\pm$0.14$^{a}$               &  GP03 \\
 M$_{B}$          & -16.92$^{b}$              &  \\
 M$_{HI}$         & 3.2$\times$10$^{8}$M$\odot$ & TM81  \\
 M$_{T}$          & 3.8$\times$10$^{8}$M$\odot$ & TM81  \\
 \hline\hline
\end{tabular}
\end{center}
\small Notes:
RA, DEC, heliocentric velocity, distance, apparent major isophotal diameter D$_{25}$ measured
at a surface brightness level of 25.0 mag~arcsec$^{-2}$, and Galactic extinction are
taken from NED (http://nedwww.ipac.caltech.edu/). The distance was
calculated using a Hubble constant of 73~km s$^{-1}$ Mpc$^{-1}$, and
taking into account the influence of the Virgo cluster, the Great
Attractor, and the Shapley supercluster.  (a) Integrated magnitudes from
\cite{GildePaz2003}, corrected for Galactic extinction; (b) absolute
magnitude in the B band computed from the tabulated B magnitude and
distance. References.- RC3: \cite{deVaucouleurs1991}; TM81: \cite{ThuanMartin1981}; 
 and GP03: \cite{GildePaz2003}.  
\end{table}

\section{The data}

\subsection{Observations and data process}

Spectrophotometric observations of Haro\,14 were performed at the Very Large Telescope (VLT;
ESO Paranal Observatory, Chile), with VIMOS in its IFU mode. They were
carried out in visitor mode, during the nights of August 19-20, 2007. The blue (HR-Blue;
4150--6200\,\AA) and orange (HR-Orange; 5250--7400\,\AA) grisms in high resolution
mode (dispersions of 0.51\,\AA\,pix$^{-1}$,  and of 0.60\,\AA\,pix$^{-1}$,
respectively) were used. A field of view (FOV) of $27\arcsec \times 27\arcsec$ on the
sky was mapped with an spatial sampling of  $0\farcs67$ (see
Figure~\ref{Figure:Haro14V-FOV} and Table~\ref{tab:data}).

Haro\,14 was observed for 4320~s with both the
HR-Blue and  HR-Orange grisms. The weather conditions were good, the seeing was in the
range 0.94-1.53 arcsec, and exposures were taken at airmass 1.03-1.23. The
spectrophotometric standard EG~274 was observed for flux calibration. 

The data were processed using the ESO VIMOS pipeline (version 2.1.11) via the
graphical user interface {\sc Gasgano}. A complete description of observations and the
data analysis has been presented in \cite{Cairos2015}.

\subsection{Creating the 2D maps}
\label{Section:mapas}

The relevant parameters of the emission lines (line flux, centroid position, line
width, and continuum) were measured by fitting single Gaussian line profiles. The fit
was carried out with the {\em Trust-region} algorithm for non-linear least squares,
using the function {\em fit} of {\sc Matlab}; reasonable initial values and lower and
upper bounds for the parameters were provided. We run an automatic procedure,
which fits a series of lines for every spaxel, namely, H$\beta$,
[\ion{O}{iii}]$\lambda4959$, [\ion{O}{iii}]$\lambda5007$, [\ion{O}{i}]$\lambda6300$,
H$\alpha$, [\ion{N}{ii}]$\lambda6583,$ and  [\ion{S}{ii}]$\lambda6716$,
[\ion{S}{ii}]$\lambda6731$. Errors of the parameters are provided by the fitting
code.

Computing the fluxes of the Balmer lines in emission is not a straightforward
task, as these fluxes can be considerably affected by the absorption of the underlying
population of stars. To take this effect into account in the H$\beta$ profile, in
which absorption wings are clearly visible,  we simply fit the line profile with two
Gaussian functions (one in emission and one in absorption) deriving in this way
absorption and emission fluxes simultaneously. In the case of H$\alpha$ the absence
of visible absorption wings makes a reliable decomposition impossible, and an alternative
approach must be adopted. We consider the equivalent width in absorption in H$\alpha$
be equal than in  H$\beta$. Assuming the same absorption equivalent width for the
Balmer lines, when the values cannot be fitted individually, has been traditionally
the most common strategy adopted \citep{McCall1985, Popescu2000}. It is a
supposition that is well founded by the predictions of evolutionary synthesis models
\citep{Olofsson1995}.

In low-surface-brightness (LSB)
regions, a spatial smoothing procedure was applied to increase the accuracy of the
fits. Depending on the signal-to-noise ratio (S/N) of the spaxel, the closest 5, 9, or
13 spaxels were averaged before the fit was carried out. In this way, we maintain the
spatial resolution of the bright regions of the galaxy, while obtaining a reasonable
S/N of the faint parts, but with a lower spatial resolution.

The  parameters of the fit were used subsequently to construct the 2D maps presented
here, taking advantage of the fact that the combined VIMOS data are arranged in a
regular 44x44 matrix.  Only spaxels with fluxes higher than 3$\sigma$ level are
considered.

Continuum maps at different spectral ranges were obtained by summing the flux
within specific wavelength intervals, selected so as to avoid strong emission
lines or residuals from the sky spectrum subtraction. Also an integrated continuum map 
was produced by summing over the whole spectral range, but
masking the spectral regions with a significant contribution of emission lines.

Line ratios maps for lines falling in the wavelength range of either grism were
simply derived by dividing the corresponding flux maps. Line ratio maps include again
only those
spaxels containing values higher than the 3$\sigma$ level.  

In the case of the H$\alpha $/H$\beta$,  the line ratio map was derived after
registering and shifting the H$\alpha$ map to spatially match the H$\beta$ map. The
amount of shift was calculated using the difference in position of the center of the
brighter \ion{H}{ii}-regions.  The shifting (about 0.27$\arcsec$ and 0.20$\arcsec$ in
{\sc ra} and  {\sc dec}, respectively) was applied using a bilinear interpolation. In
order to correct for the fact that  H$\alpha$ and H$\beta$ had been observed under
different seeing conditions,  the H$\alpha$ map has been convolved with a Gaussian
profile, to match both point spread functions.

All the maps present in this work were corrected for interstellar extinction in terms of
spaxels. The method is described in Section~\ref{Section:reddening} for integrated spectra; in the
maps, the same method was applied to every spaxel.

\section{Results}
\label{Section:analysis}

\begin{figure}
\centering
\includegraphics[angle=0, width=0.8\linewidth]{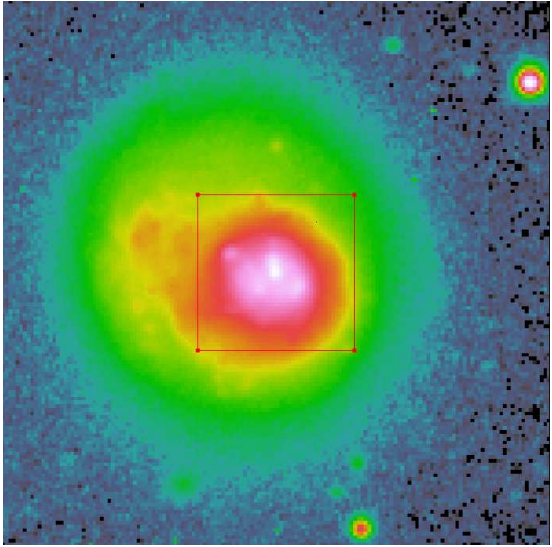}
\caption{V-band image of Haro\,14 retrieved from NED and published in
\cite{HunterElmegreen2006};
the field of view is $1.5 \arcmin$ and the central box
indicates the $27\arcsec \times 27\arcsec$ covered by VIMOS. North is up and east to the left (also
in all maps shown from here on).}
\label{Figure:Haro14V-FOV} 
\end{figure}

Haro\,14 is made of an extended high-surface-brightness (HSB) star-forming region, placed
atop a smooth LSB underlying  stellar component
(Figure~\ref{Figure:Haro14V-FOV}); this redder population has a diameter of about 4.12 kpc
\citep{deVaucouleurs1991}, whereas the central HSB region, which has been resolved in
several \ion{H}{ii}-regions and/or stellar clusters \citep{Doublier1999, Noeske2003,
Cairos2015}, fills the VIMOS FOV (about 1.7$\times$1.7~kpc). According to the morphological
classification made by  \cite{Loose1987}, the galaxy belongs to the most common BCG type,
namely, the iE objects, which represent about 70\% of the BCG population.

\subsection{Stellar and ionized gas morphology}

\label{Section:morphology}

\begin{figure*}
\centering
\includegraphics[angle=0, width=0.9\linewidth]{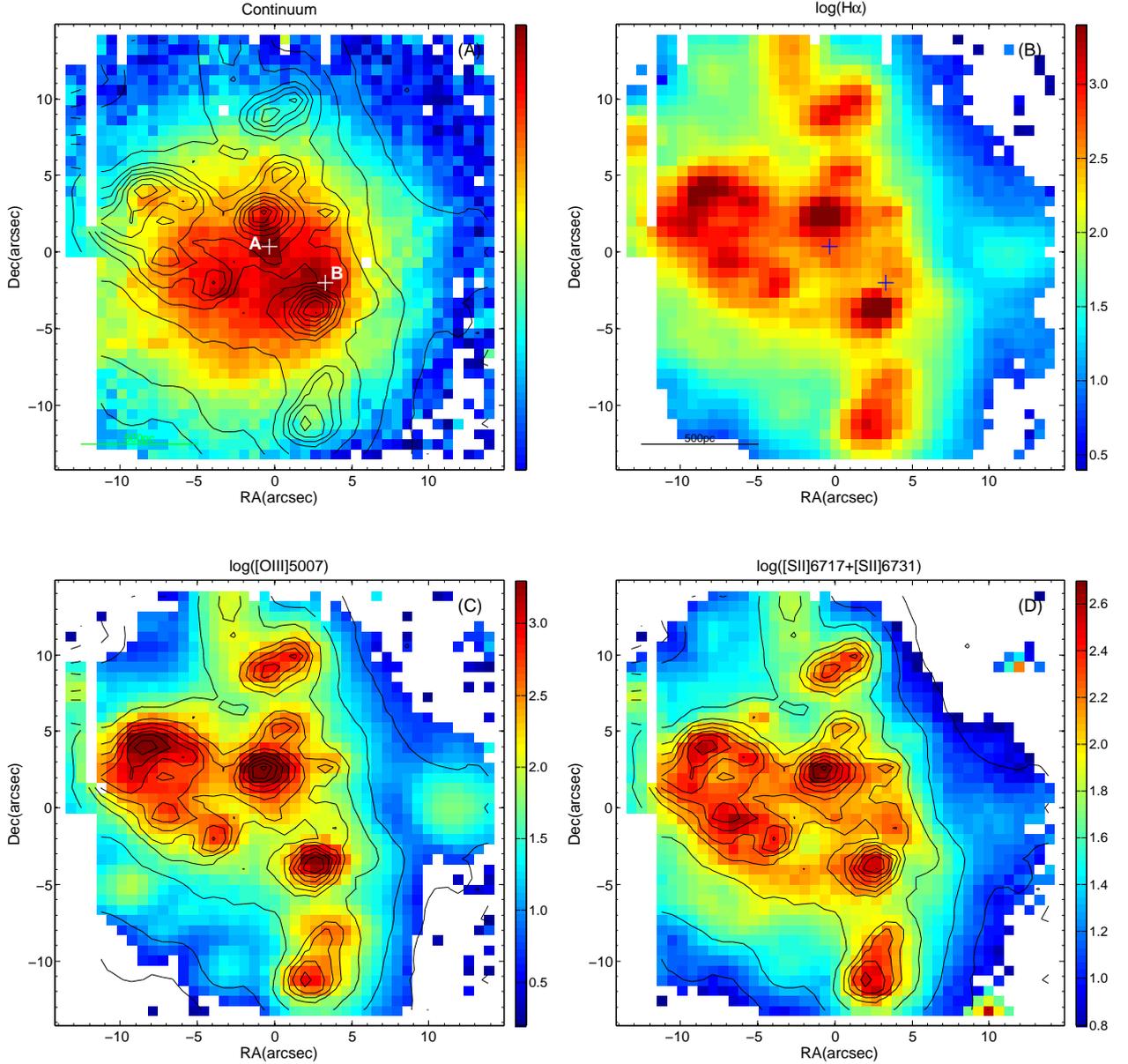}
\caption{Panel A: Haro\,14 intensity distribution in continuum with H$\alpha$ contours overplotted; the map has been obtained by summing over the whole
orange spectral range, but masking the emission lines. The two major continuum 
clusters are labeled {\sc a} and {\sc b}. The scale in parsec is indicated at the
bottom left. The image is scaled in arbitrary flux units. Panel B: H$\alpha$ 
emission-line flux (flux units are $10^{-18}$\,erg\,s$^{-1}$\,cm$^{-2}$). Panel C:
[\ion{O}{iii}]~$\lambda$5007 emission-line flux with H$\alpha$ contours overplotted. Panel~D: [\ion{S}{ii}]~$\lambda\lambda$~6717,6731 
emission-line flux  with H$\alpha$ contours overplotted.}
\label{Figure:HaroCont} 
\end{figure*}

Continuum and emission-line intensity maps of Haro\,14, derived from the VIMOS data  as
described in Section~\ref{Section:mapas}, are shown in Figure~\ref{Figure:HaroCont}. At the
galaxy distance, the resolution element (spaxel) translates into 42 pc.

The continuum registers photons produced in the photosphere of the stars and, hence,
traces the stellar component of the galaxy. On the other hand, hydrogen
recombination-line maps capture photons produced by the excited atoms of the gas, as
a recombined electron cascades down to the ground state. Forbidden lines of trace
species, such as [\ion{O}{iii}] or [\ion{N}{ii}], the other main type of emission-lines
detected in the spectra of BCGs/\ion{H}{ii}-galaxies, are also generated in the
ionized interstellar gas; these lines result from the emission of photons from
collisionally excited levels. Therefore, emission-line maps trace regions of
photoionized hydrogen, i.e., regions that are rich in hot OB stars.

\begin{figure}
\centering
\includegraphics[angle=0, width=0.9\linewidth]{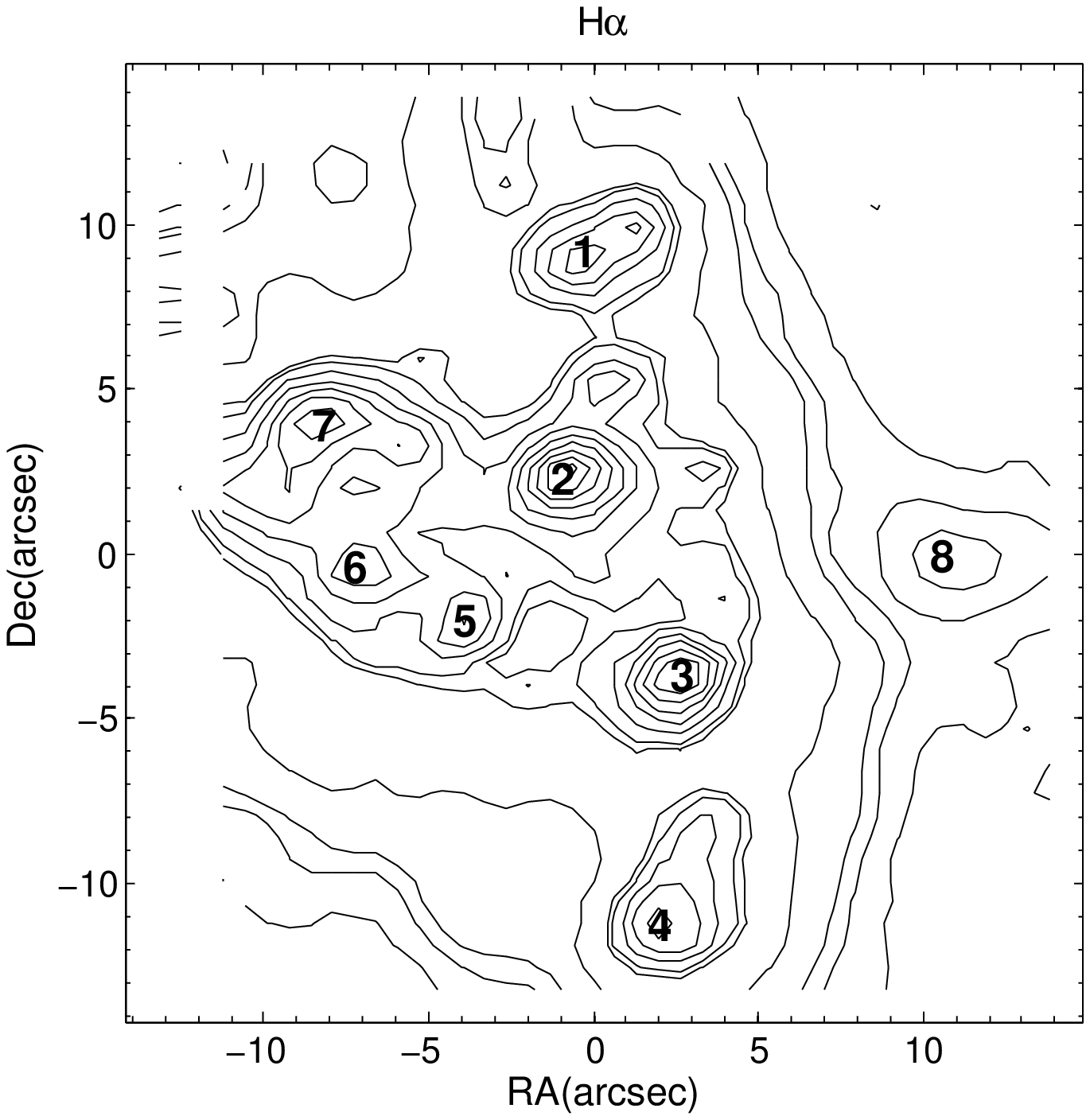}
\caption{Contour map of the H$\alpha$ emission in Haro\,14 with the major 
regions of SF labeled.}
\label{Figure:Ha-contour} 
\end{figure}

Substantial deviations between the stellar and ionized gas distributions in Haro\,14
are evident from Figure~\ref{Figure:HaroCont}. In the continuum, the intensity
reaches its maxima close to the galaxy center, where an ensemble of central stellar
clusters is identified, encircled by a roughly regular stellar envelope. The two major
clusters were labeled as {\sc a, b} following \cite{Noeske2003}. The maximum of
emission in the continuum, which does not depend on the spectral window selected, is
cospatial with Knot~{\sc a}.

The maps in warm ionized gas exhibit a quite different morphology.  Here, the peaks of intensity do
not gather in the central galaxy regions, but they are scattered across the whole mapped area. A
linear (chain-like) structure extends in the north-south direction, and a horseshoe-like curvilinear
feature departs from these central knots and extends about $12\farcs0$ (760~pc) eastward; the
morphology and size of this curvilinear feature suggest the presence of a large-scale expanding
structure (supershell or superbubble; \citealp{TenorioTagle1988}). The major SF knots are labeled in
Figure~\ref{Figure:Ha-contour}. 

The central galaxy region displays essentially the same pattern in all the
emission-line maps with the intensity maximum always situated in Knot~{\sc 2} (see
\citealp{Cairos2015} for a complete collection of emission-line maps for Haro\,14).
Extended diffuse gas emission surrounds these major \ion{H}{ii} regions. Several
fainter lumps emerge around the central SF major area: the largest lump at the west 
(Knot~8 in Figure~\ref{Figure:Ha-contour}) and some others detectable in the
southeast. These blobs are better seen in the higher excitation 
[\ion{O}{iii}] line, where they are well visible as condensations separated from the
galaxy main body, whereas  they are almost undetectable in the [\ion{S}{ii}] lower
ionization line (see Figure~\ref{Figure:HaroCont}).

None of the bright continuum peaks (Knots~{\sc a} and {\sc b}) spatially
coincide with a peak in the ionized-gas distribution, indicating that the HSB
clusters are dominated by non-ionizing stars (ages larger than 10 Myr). The
continuum peak (Knot~{\sc a}) is located about 168~pc south of the emission-line
peak (Knot~2). 

In order to understand the star-forming history of a galaxy, its different stellar
components must be first distinguished and characterized. By comparing the
emission-line and continuum morphology in Haro\,14, we get the first insights on its
stellar content: at least three different populations are identified in this galaxy. We disentangled two distinct 
episodes of SF in the central region: the
very young ionizing stars, i.e., Knots~1-8  (ages $\leq$ 10 Myr), and an
intermediate-age population (the HSB clusters visible in the continuum frames,
Knots~{\sc a} and {\sc b}). Both of these stellar components rest on an underlying LSB host galaxy,
whose optical and NIR colors indicate an age of several Gyr \citep{Marlowe1999,Noeske2003}. 
The properties
of the two more recent episodes of SF identified in the galaxy  
are addressed in the next sections.

\subsection{Extinction and its distribution}

\begin{figure}
\centering
\includegraphics[angle=0, width=0.9\linewidth]{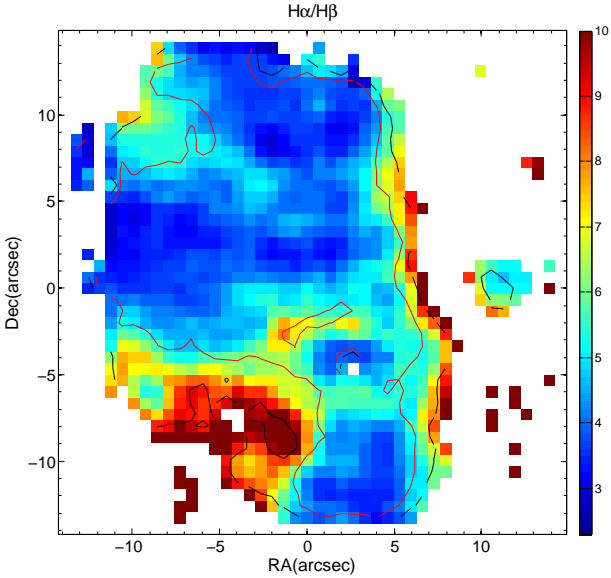}
\caption{H$\alpha$/H$\beta$ emission-line ratio for Haro\,14. The map contains 
only spaxels with
values higher than 3 $\sigma$ level. 
Black and red
lines delineates the isocontours at 5 $\sigma$ level and 10 $\sigma$ level, 
respectively.}
\label{Figure:reddening} 
\end{figure}

In the optical domain, the interstellar extinction is derived from the ratio of the different
\ion{H}{i} Balmer line series to H$\beta$.  The use of IFS data allows us to compute the
distribution of the dust across the whole observed FOV.

Figure~\ref{Figure:reddening} shows the H$\alpha$/H$\beta$ line-ratio map of 
Haro\,14. The galaxy  presents a highly inhomogeneous extinction pattern: the areas
of SF are practically dust-free regions (H$\alpha$/H$\beta\lesssim 3$), whereas the
ratio H$\alpha$/H$\beta$ considerable increases when moving toward the periphery of
the knots. The largest concentrations of dust are found in the zones of lowest
H$\alpha$ emission, namely, the southeast and western regions. That is consistent
with a scenario in which dust is destroyed or swept by the most massive stars.

The extinction coefficient, C(H$\beta$), for every individual spaxel can be 
derived from the ratio H$\alpha$/H$\beta$ (see Section~\ref{Section:reddening}).
Typical values of H$\alpha$/H$\beta$ close to the center of the SF knots are about 3, whereas in
the galaxy periphery values of 10 are easily reached. Such ratios imply extinction
coefficients C(H$\beta$)=0.06 and 1.57, respectively, which translate into
magnitudes of extinction in $V$ of A$_{V}$=0.12 and A$_{V}$=3.38, and
color excess of $E(B-V)$=0.04 and $E(B-V)$=1.09\footnote{with A$_{V}$=2.16$\times$C(H$\beta$) and $E(B-V)$=0.697$\times$C(H$\beta$);
\citealp{Dopita2003}.}.  Such large variations in $E(B-V)$ stress the importance of
obtaining two-dimensional information on the dust distribution even when we are
dealing with small and low-metal content systems such as dwarfs. The use of an unique
extinction coefficient for the whole galaxy (as usual when the spectroscopic
data are taken using a slit) can yield large errors in the extinction correction
and, hence, in the derived fluxes and magnitudes.

Besides dust, other factors can affect the H$\alpha$/H$\beta$ ratio. In particular,
a large component of collisional excitation to the hydrogen lines can enhance
H$\alpha$/H$\beta$. That collisional excitation could  play a major role in
increasing the H$\alpha$/H$\beta$ in the southeastern and western galaxy regions is
supported by the fact that the areas of high H$\alpha$/H$\beta$ coincide
very well
with the regions where shocks are the dominant ionization mechanism (see
Section~\ref{ionization} and Figure~\ref{Figure:map-diagnostic}).

\subsection{The electron density distribution}

\begin{figure}
\centering
\includegraphics[angle=0, width=0.9\linewidth]{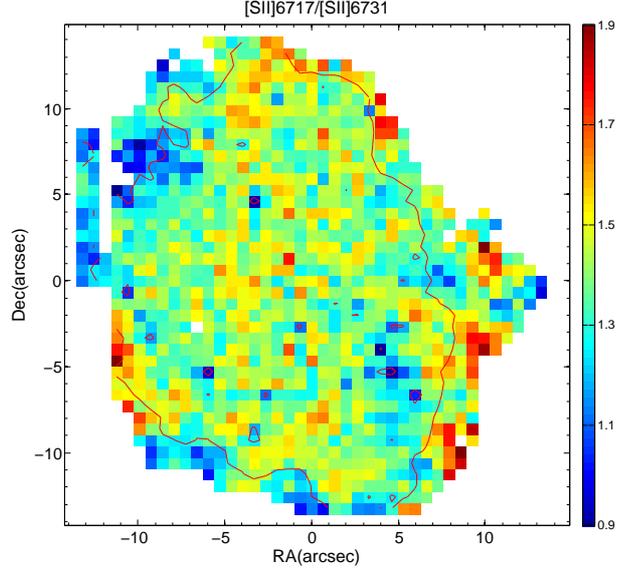}
\caption{Electron density sensitive
[\ion{S}{ii}]~$\lambda6717$/[\ion{S}{ii}]~$\lambda6731$ emission-line ratio. The map contains 
only spaxels with
values larger than 3$\sigma$-level. The red
line delineates the isocontour at 5$\sigma$-level.}
\label{Figure:density} 
\end{figure}

The electron density, N${e}$, can be derived by observing the effects of collisional
deexcitation; the most often used line ratio is
[\ion{S}{ii}]~$\lambda6717$/[\ion{S}{ii}]~$\lambda6731$, a sensitive density diagnostic
in the range 100 to 10000~cm$^{-3}$. 

The central area of the map of 
[\ion{S}{ii}]~$\lambda6717$/[\ion{S}{ii}]~$\lambda6731$ 
(Figure~\ref{Figure:density}) shows only small fluctuations of the line ratio, between
1.5 and 1.2, which at T=10000~K correspond to density in the range 
$\leq$100~cm$^{-3}$ up to 240cm$^{-3}$. The electron density increases slightly at
the galaxy northeast and southwest regions, where values  of the line ratio of
0.90$\pm$0.16 translates into densities of about 900~cm$^{-3}$.

As expected, the SF knots show mostly low density values (N$_{e}\leq$100cm$^{-3}$),
whereas the density peaks are reached in the periphery of the knots ($\geq$ 180-200~pc).
Areas of higher density at the outskirts of the SF regions are most probably associated
with filaments, which are formed when the expanding shell front hits the ISM.

\subsection{On the ionizing mechanism acting in Haro\,14}
\label{ionization}

\begin{figure}
\centering
\includegraphics[angle=0, width=0.9\linewidth]{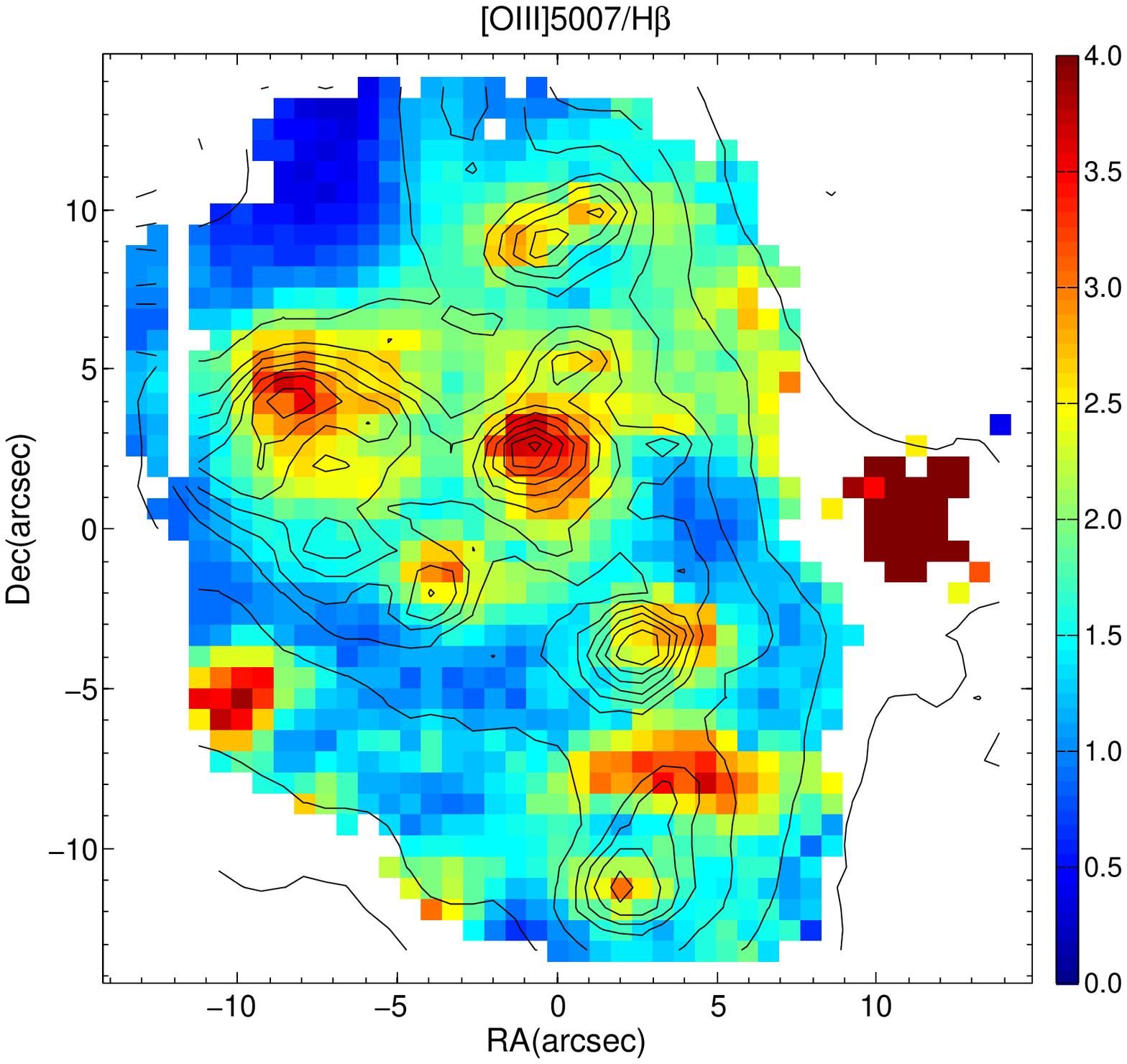}
\caption{[\ion{O}{iii}]~$\lambda5007$/\Hb{} emission-line ratio map with contours on 
H$\alpha$ overplotted.}
\label{Figure:diagnostic-oiii} 
\end{figure}

\begin{figure}
\centering
\includegraphics[angle=0, width=0.9\linewidth]{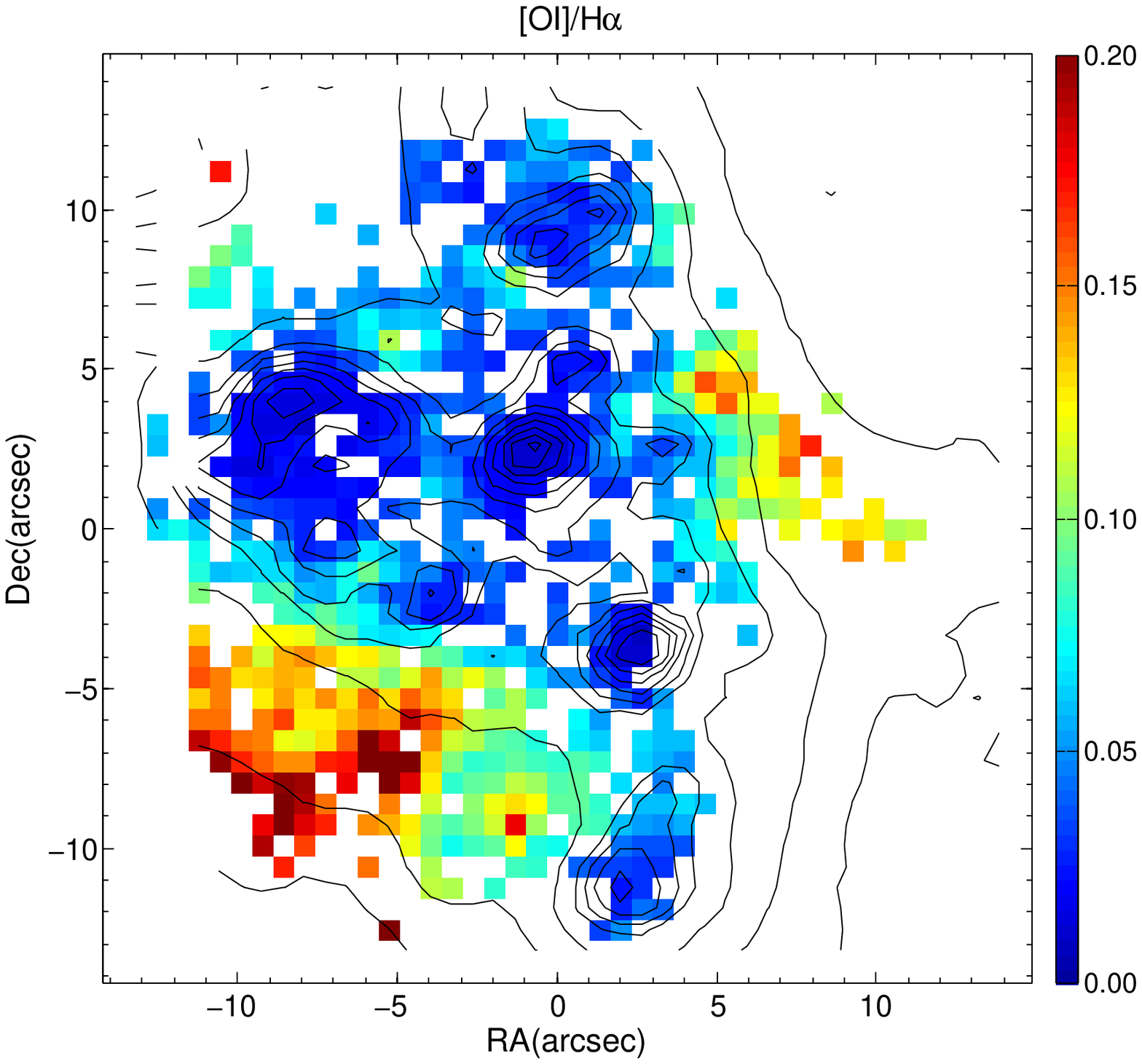}
\caption{[\ion{O}{i}]~$\lambda6300$/\Ha{} emission-line ratio map with contours on 
H$\alpha$ overplotted.}
\label{Figure:diagnostic-oi} 
\end{figure}

\begin{figure}
\centering
\includegraphics[angle=0, width=0.9\linewidth]{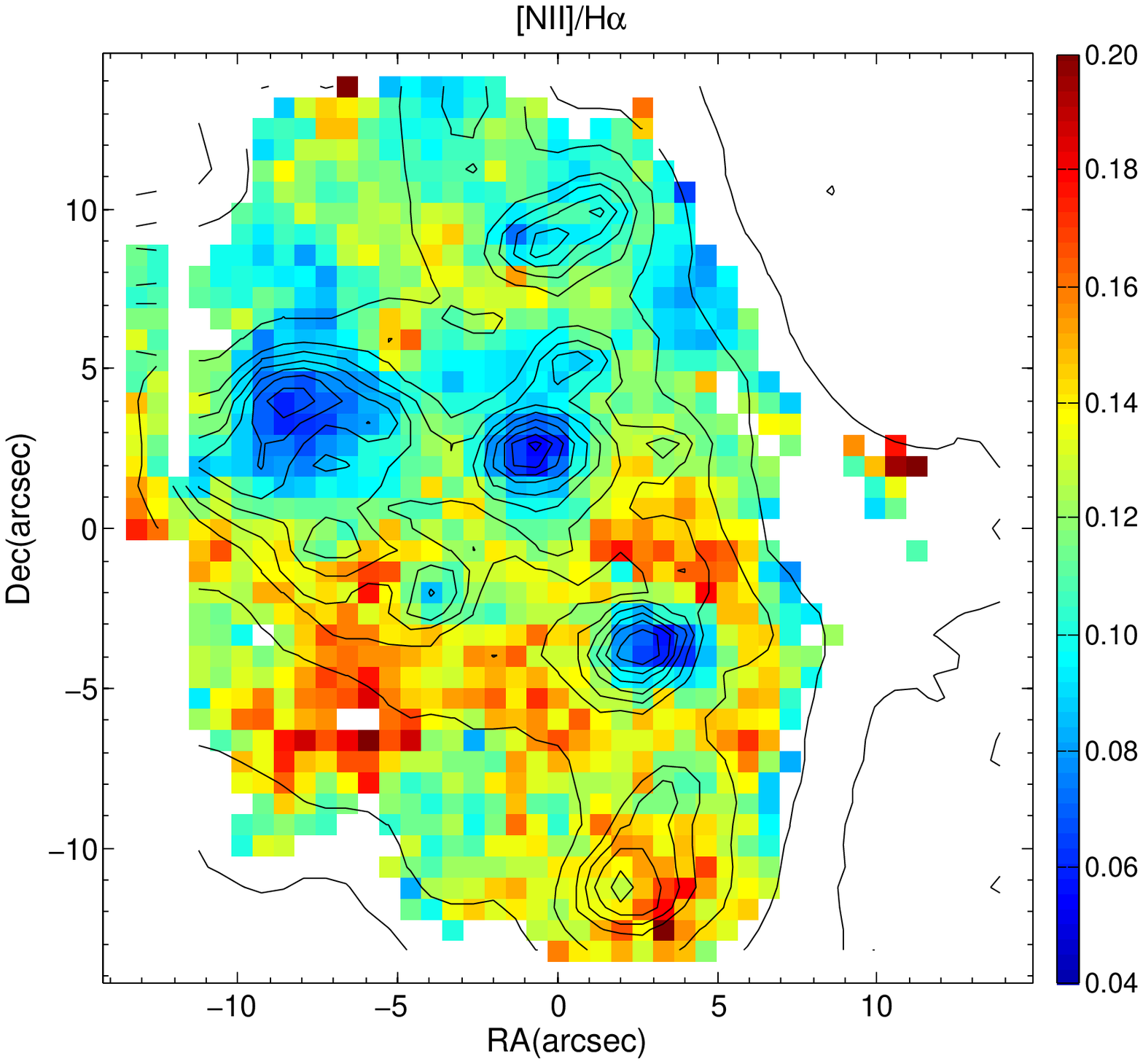}
\caption{[\ion{N}{ii}]~$\lambda6584$/\Ha\ emission-line ratio map with contours on 
H$\alpha$ overplotted.}
\label{Figure:diagnostic-nii} 
\end{figure}

\begin{figure}
\centering
\includegraphics[angle=0, width=0.9\linewidth]{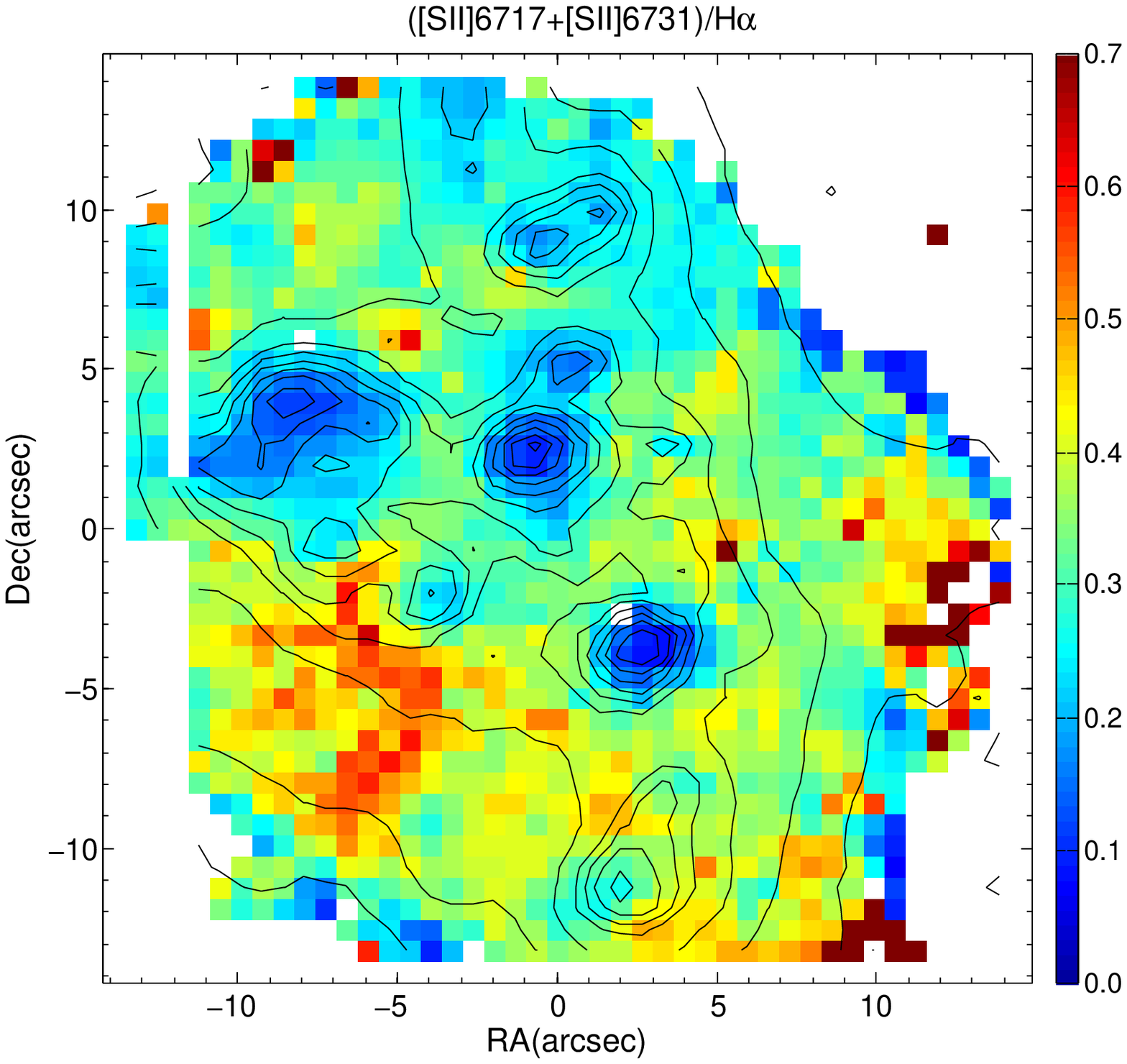}
\caption{
[\ion{S}{ii}]~$\lambda\lambda6717,\,6731$/\Ha{} emission-line ratio map with contours on 
H$\alpha$ overplotted.}
\label{Figure:diagnostic-sii} 
\end{figure}

Two main mechanisms are responsible for the gas ionization in a nebula:
\emph{photoionization}, which is the ionization due to high-energetics photons produced
in hot stars or in active galaxy nuclei (AGN), and \emph{shock-ionization}, which is
the collisional ionization that takes place in shocks caused by stellar winds and
supernovae \citep{Dopita2003}. 

Extinction-independent line intensity ratios are commonly used to investigate the
ionization mechanism  in emission-line galaxies \citep{Baldwin1981,Veilleux1987}. The 
diagnostic line ratios most frequently employed are [\ion{O}{iii}]~$\lambda5007$/\Hb{},
[\ion{N}{ii}]~$\lambda6584$/\Ha, [\ion{S}{ii}]~$\lambda\lambda6717,\,6731$/\Ha{} and 
[\ion{O}{i}]~$\lambda6300$/\Ha{}.

The two-dimensional information provided by the IFS data allows us to investigate the
ionization mechanism acting in Haro\,14 across the whole area covered with VIMOS.

The [\ion{O}{iii}]~$\lambda5007$/\Hb{} map of Haro\,14 displays an intriguing pattern
(see Figure~\ref{Figure:diagnostic-oiii}). Since [\ion{O}{iii}]$\lambda5007$ arises from an ion of
relatively high ionization potential (35.1~eV), [\ion{O}{iii}]~$\lambda5007$/\Hb{}  is
a good indicator of the radiation field strength and, as expected, we find  that the
major SF knots  are associated with peaks in  [\ion{O}{iii}]~$\lambda5007$/\Hb{}.
However, there are also several other regions in the galaxy, which are not associated with any of
the identified  \ion{H}{ii} regions, with  very high-excitation values; indeed, the
largest excitation (values up to 6) is reached in the Knot~8, the faint blob situated at the west of the galaxy main body. High values of the
excitation ([\ion{O}{iii}]~$\lambda5007$/\Hb{} $\approx$ 4) are also found in a clump
detected at the galaxy southeast area and,  even more puzzling, a patch of high
excitation (up to 3.7) is found in the region between Knots~3 and 4 (see 
Figure~\ref{Figure:diagnostic-oiii}).

Several factors can affect the [\ion{O}{iii}]~$\lambda5007$/\Hb{} ratio. High values of
[\ion{O}{iii}]~$\lambda5007$ imply a high ionization parameter. Hence in regions photoionized
by OB stars, higher excitations are indicative of very high temperatures, and also very hot (very
young) OB stars. Alternatively, another ionization mechanism, as AGN or shocks, can be
responsible for the high excitation.

\begin{figure*}
\centering
\includegraphics[angle=0, width=0.9\linewidth]{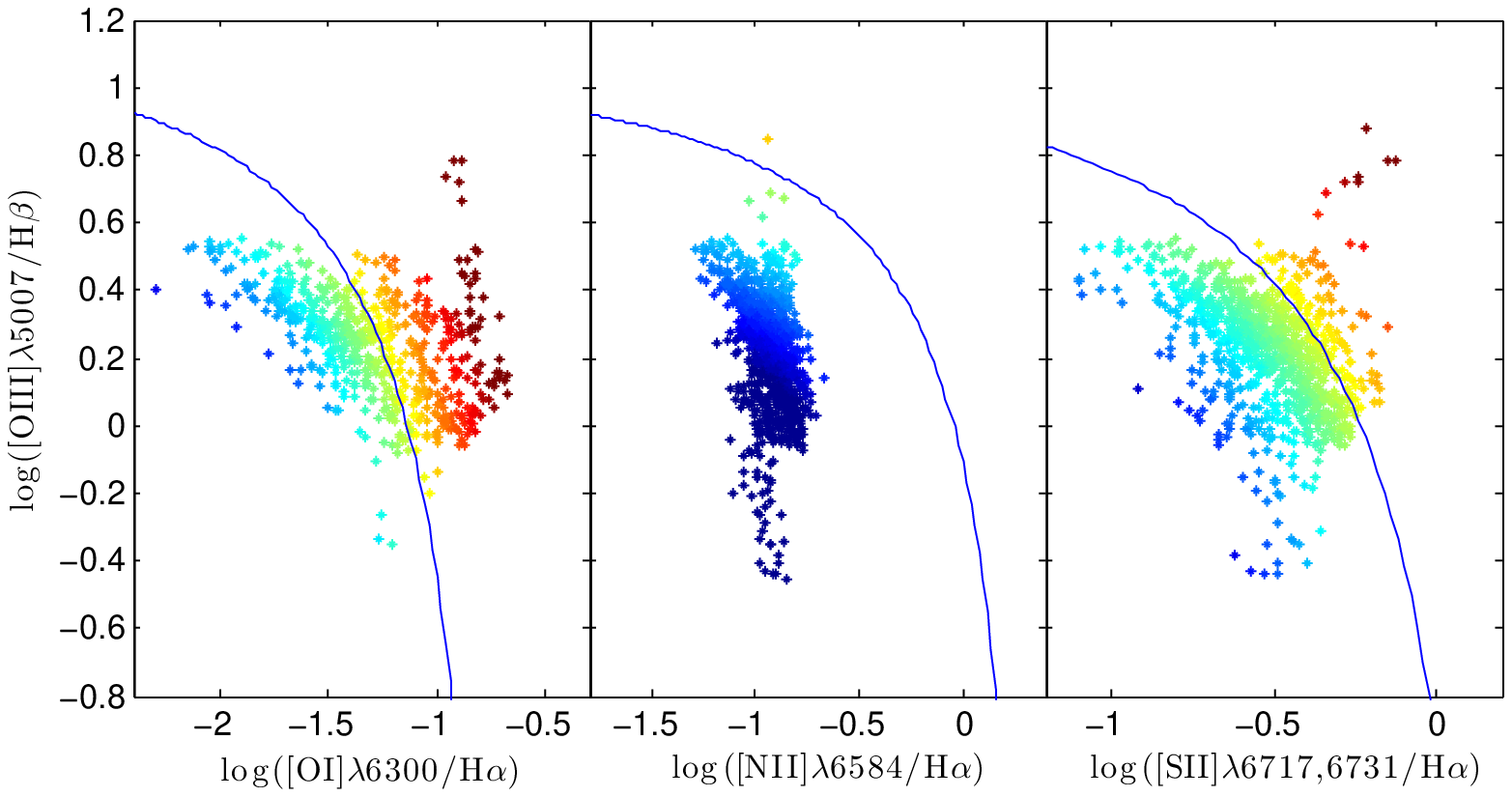}
\caption{Optical emission-line diagnostic
diagram for the different spaxels in Haro\,14. The theoretical maximum starburst line  
derived by \cite{Kewley2001} is also
included in the figure. To better visualize the results on the diagram, the points were
color coded according to their distance to the maximum starburst line. In all three diagrams, the lower left section
of the plot is occupied by spaxels in which the dominant energy source is radiation from hot stars (blue points in the figure). Additional ionizing mechanisms
shift the spaxels to the top right and right part of the diagrams (yellow to red colors). 
}
\label{Figure:diagnostic-spaxel} 
\end{figure*}

\begin{figure*}
\centering
\includegraphics[angle=0, width=0.8\linewidth]{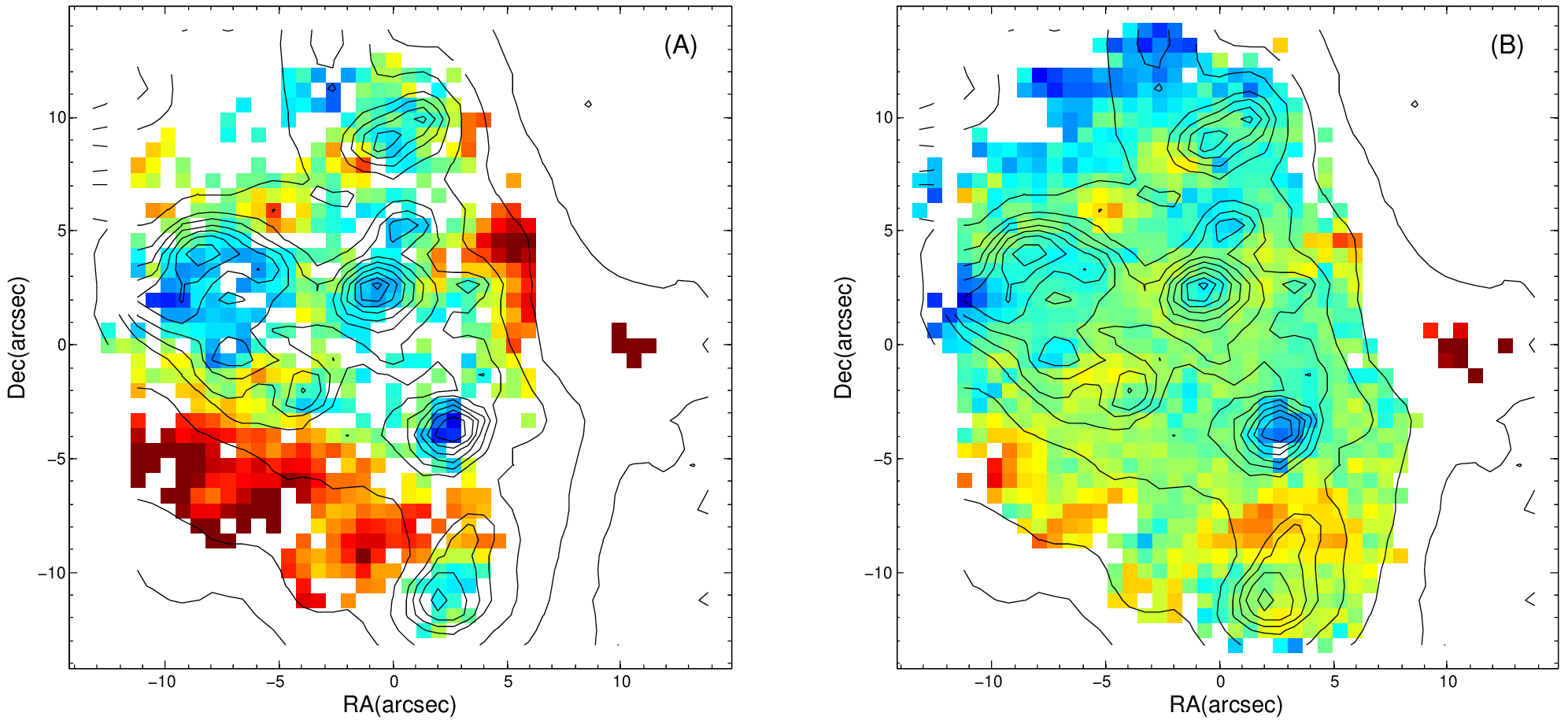}
\caption{Spatial localization of the spaxels in the diagnostic diagrams [\ion{O}{iii}]~$\lambda5007$/\Hb{} vs. 
[\ion{O}{i}]~$\lambda6300$/\Ha, and [\ion{O}{iii}]~$\lambda5007$/\Hb{} vs.
[\ion{S}{ii}]~$\lambda\lambda6717,\,6731$/\Ha{}. The color code is the same 
as in Figure~\ref{Figure:diagnostic-spaxel}. }
\label{Figure:map-diagnostic} 
\end{figure*}

The three high-excitation zones in Haro\,14, which do not coincide  spatially with  
bright star-forming regions, are (obviously) not powered by an AGN. Although a
population of very hot and massive (young) stars can  contribute to increase the
excitation, Figure~\ref{Figure:diagnostic-spaxel} and Figure~\ref{Figure:map-diagnostic}
point to shocks   as the most plausible origin of the hard radiation able to produce
such  high-excitation values.

The [\ion{O}{i}]~$\lambda6300$/\Ha{}, [\ion{N}{ii}]~$\lambda6584$/\Ha,\ and
[\ion{S}{ii}]~$\lambda\lambda6717,\,6731$/\Ha{} maps are shown in 
Figures~\ref{Figure:diagnostic-oi}-\ref{Figure:diagnostic-sii}. The three maps represent
 a similar pattern; they
trace the regions of SF and present low values in the center of the knots, 
which increase outward.  In the galaxy outskirts,  both
[\ion{O}{i}]~$\lambda6300$/\Ha{} and [\ion{S}{ii}]~$\lambda\lambda6717,\,6731$/\Ha{}
reach values that are inconsistent with photoionization by hot stars
([\ion{O}{i}]~$\lambda6300$/\Ha{} $\geq$ 0.1 and
[\ion{S}{ii}]~$\lambda\lambda6717,\,6731$/\Ha{} $\geq$ 0.5), indicating that another
ionizing mechanism, most probably  shocks, is acting there.

To better distinguish among different ionization sources, diagnostic-diagrams are commonly
employed \citep{Veilleux1987,Kewley2001, Kauffmann2003, Kewley2006}. In combination with
two-dimensional spectroscopy, this allows us to probe the power sources acting in different
galaxy regions by plotting the spectra from each individual spaxel in the diagnostic
diagrams. In recent years this technique has been successfully applied to IFU data
\citep{Sharp2010,Rich2011,Rich2012,Rich2015,Leslie2014,Belfiore2015,Belfiore2016}.

Figure~\ref{Figure:diagnostic-spaxel} shows the diagnostic-line ratios for the individual spaxels 
in Haro\,14 on the most widely employed diagnostic diagrams, namely,
[\ion{O}{iii}]~$\lambda5007$/\Hb{} versus  [\ion{O}{i}]~$\lambda6300$/\Ha,
[\ion{N}{ii}]~$\lambda6584$/\Ha, and  [\ion{S}{ii}]~$\lambda\lambda6717,\,6731$/\Ha. The maximum
starburst line or photoionization line from  \cite{Kewley2001} is also shown in the figure. 
This boundary indicates the limit between gas photoionized by hot stars and gas ionized via other
mechanisms; the flux ratios of any object lying above this boundary cannot be modeled by hot stars
photoionization, but require an additional contribution from a harder radiation source such as an AGN
or shock excitation.

In the [\ion{O}{iii}]~$\lambda5007$/\Hb{} versus
[\ion{S}{ii}]~$\lambda\lambda6717,\,6731$/\Ha, and [\ion{O}{iii}]~$\lambda5007$/\Hb{}
versus[\ion{O}{i}]~$\lambda6300$/\Ha\  diagrams a significant number of spaxels fall 
out of the areas occupied by photoionization for stars. 
The ratio [\ion{N}{ii}]~$\lambda6584$/\Ha \ , which is only weakly dependent on
the hardness of the radiation field, but strongly dependent on the metallicity, is
not effective at separating shocks from photoionized gas \citep{Allen2008,Hong2013}.

Figure~\ref{Figure:map-diagnostic} shows the spatial location in the galaxy of the points plotted
in the diagnostic diagrams; the maps are color coded as in Figure~\ref{Figure:diagnostic-spaxel},
as the redder regions are those that lie further from the maximum starburst line from
\cite{Kewley2001}. The non-photoionized areas are situated mainly in the galaxy outskirts. In
particular, the    southwest and west regions lie far above the maximum starburst line, and
non-photoionized arc and circular features are visible around the major SF regions. Such an arc of
non-photoionized gas, previously seen in star-forming galaxies, can be explained as the result of
the interaction of two (or more) winds, originated in different stellar clusters.  This pattern
allows us to conclude with safety that shocks, generated  by the mechanical feedback of stellar
winds and supernovae explosions, are acting in Haro\,14.

\subsection{Integrated spectroscopy}

We identify and delimit the major \ion{H}{ii} regions and the likely young stellar
clusters in Haro\,14. Then, we add the spectra of the spaxels in each region to produce
a higher S/N spectrum from which we determine reliable physical
parameters and abundances. 

\ion{H}{ii} regions are very well traced in \ion{H}{i} recombination-line maps, in
which the emission is produced by excited hydrogen atoms decaying to lower levels by
radiative transitions. We outline the \ion{H}{ii} regions in Haro\,14 by hand from 
the H$\alpha$ maps: this line has a higher S/N and is not heavily affected by stellar
absorption compared with H$\gamma$ or H$\beta$. Young stellar clusters, dominated
by hot and blue, but non-ionizing stars, are well visible as HSB regions in continuum
maps or bluer regions in color maps. To isolate the young stellar clusters in
Haro\,14, we made use of the continuum maps. There is not a clear cut criteria to
define the limits of these regions. We adopted thence the following approach: we
integrated over a boundary that traces the morphology of the clusters, keeping in
mind that the minimum size is limited by the seeing; being conservative, we
considered only regions with more than 10 spaxels.

We generated integrated spectra of the eight knots identified in the H$\alpha$ maps (Knots~1
to 8 in Figure~\ref{Figure:Ha-contour}), and of the two major HSB stellar clusters seen in
continuum (Knots~{\sc a} and {\sc b}). The main properties of the individual spectra are
highlighted below.

\begin{figure*}
\centering
\includegraphics[angle=0, width=1\linewidth]{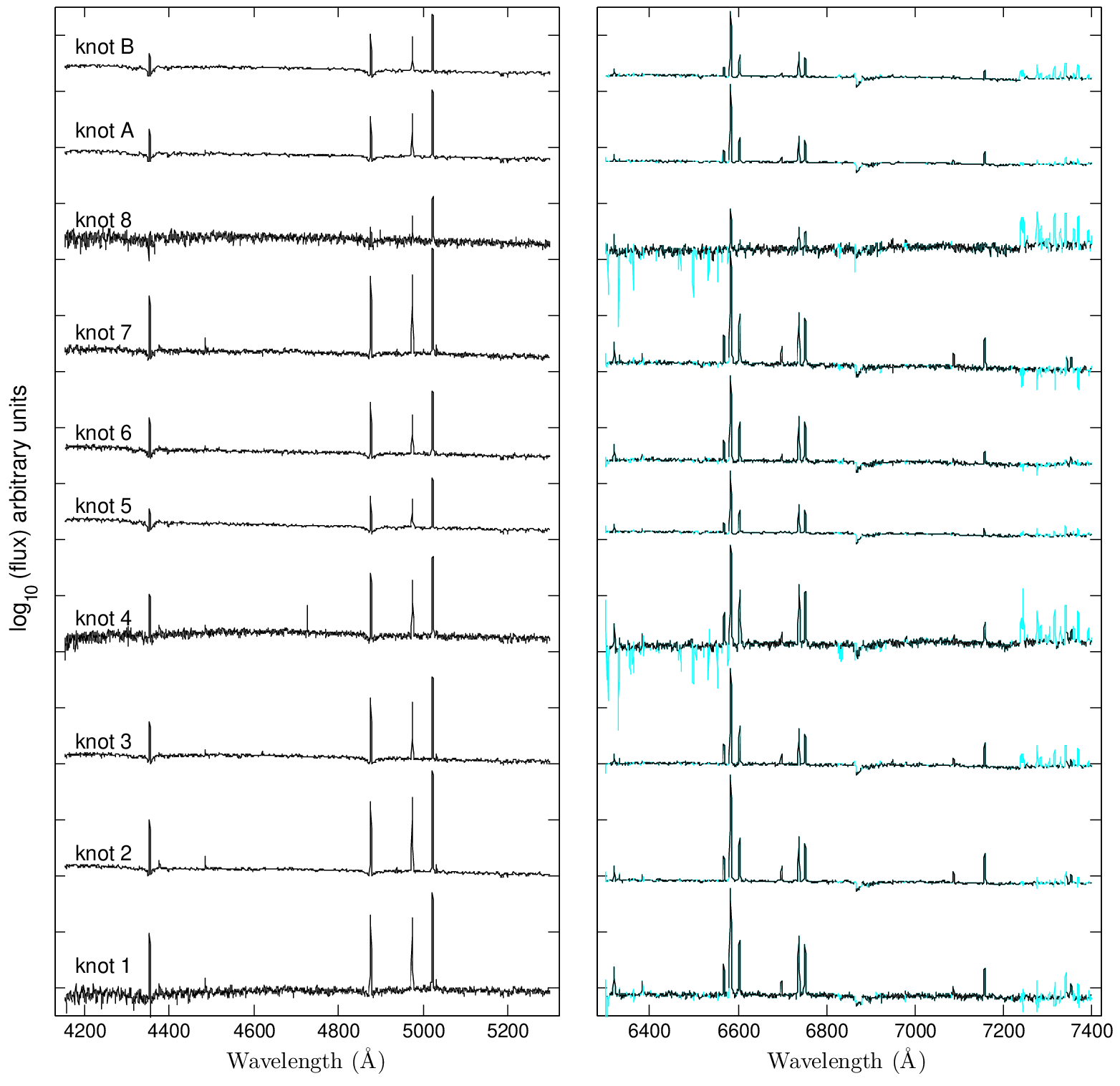}
\caption{Flux calibrated spectra of the eight selected \ion{H}{ii} regions and the
two HSB stellar clusters in Haro\,14, 
in logarithmic units. The regions of the spectra affected by the sky lines are 
plotted in cyan.}
\label{Figure:Haro14spec} 
\end{figure*}

\begin{figure*}
\centering
\includegraphics[angle=0, width=1\linewidth]{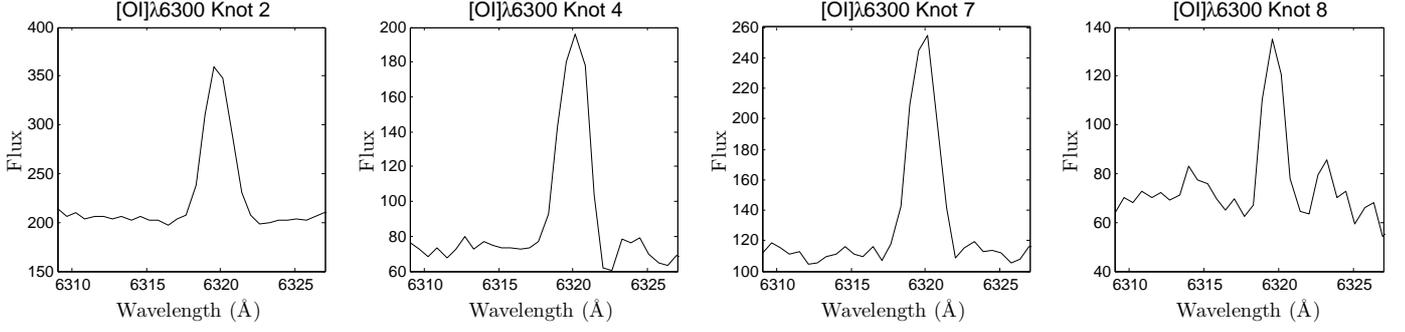}
\caption{[\ion{O}{i}]~$\lambda6300$ line profile  in four of the selected \ion{H}{ii}
 regions in Haro~14, including the two regions most heavily affected by sky
 residuals, knots~4 and 8. None of the residuals fall above the line, and we can
 reliably measure its flux.}
\label{Figure:Haro14-oi-line} 
\end{figure*}

All knots show \ion{H}{ii} regions spectra with strong emission features atop an almost flat
continuum, but absorption features around H$\gamma$ and H$\beta$ are also seen  (see 
Figure~\ref{Figure:Haro14spec}). Residuals from the sky subtraction are evident in several
spectra; as explained in \cite{Cairos2015}, the subtraction of the sky was a difficult part
of the data process. However, these residuals constitute mostly a cosmetic
problem because they are located in spectral regions where important features do not fall;
mainly the region around [\ion{O}{i}]~$\lambda6300$ and the
spectral region above 7200\AA\ in knots~4 and 8 are affected by sky residuals. Fortunately, the line
[\ion{O}{i}]~$\lambda6300$, redshifted to 6320\AA\ ,  can also be accurately measured in knots~4
and 8.  Figure~\ref{Figure:Haro14-oi-line} shows the [\ion{O}{i}]~$\lambda6300$ profile in
four of the spectra, including the two more problematic, Knots~4 and 8. It is clear that the line
can be reliably measured. Above 7200\AA\ no lines have been measured in these two knots.

$\bullet$ {\em Knot~1} belongs to the SF chain structure, where it is located about
590~pc to the north of Knot~{\sc a}. It presents a typical nebular spectrum, with
prominent Balmer emission lines and  strong [\ion{O}{iii}],  [\ion{N}{ii}], and 
[\ion{S}{ii}] forbidden lines atop a weak and rather flat continuum. The flatness
of the spectrum and the lack of pronounced absorption features indicates a 
significant contribution of the nebular continuum.

$\bullet$ {\em Knot~2} is the region where the emission lines are strongest. The
conspicuous increase of the continuum toward the blue and the presence of strong
absorption wings in the higher terms of the Balmer series, H$\gamma$ and H$\beta$,
implies a sizable stellar contribution---most probably of hot OB and A stars.

$\bullet$ {\em Knot~3} also belongs to the SF central chain. Its spectrum  steps
up markedly toward the blue and the presence of an
intermediate-age stellar population to the continuum is evidenced by the
absorption in the higher order hydrogen Balmer-lines - most probably
starlight coming from Knot~{\sc b}, which is displaced only about 130~pc
northwest.

$\bullet$ {\em Knot~4}, the southeast knot in the central chain structure, is
displaced about 725~pc southwest from the continuum peak. Its spectrum reveals
strong emission lines on top of an almost flat continuum, in which small
absorption wings are in view around H$\gamma$ and H$\delta$ in emission.

$\bullet$ {\em Knot~5} is a small \ion{H}{ii} region, located in the horseshoe-like
curvilinear structure. The spectrum shows a steep continuum, raising toward the
blue and strong absorption on the Balmer lines;  absorption lines, such as
CaI~$\lambda$4226 or MgII~$\lambda$4481, are clearly distinguished in the spectrum and are indicative of a significant contribution
from intermediate-age stars.  

$\bullet$ {\em Knot~6} is also a small \ion{H}{ii} region, located in the loop.  As in
the case of knot~5, the spectrum increases notably toward the blue,  shows strong
absorption wings on the Balmer lines and absorption features (CaI~$\lambda$4226 or
FeI~$\lambda$4383) characteristic of A stars.

$\bullet$ {\em Knot~7} is the largest \ion{H}{ii} region in the horseshoe-like
curvilinear structure; it represents an almost flat spectrum, which increases
slightly toward the blue and is dominated by the emission lines of the ionized gas. 

$\bullet$ {\em Knot~8} is located on the west, off the galaxy main body. It is a quite
faint blob, most clearly seen in the high-excitation [\ion{O}{iii}] line.  It presents
a flat spectrum; the poor S/N does not allow us to resolve any other feature except the
brightest emission-lines.

$\bullet$ {\em Knot~A} is the peak in continuum. The emission increases visibly toward
the blue, and strong absorption wings in the higher order Balmer recombination lines
and several absorption features, such as CaI~$\lambda$4226, FeI~$\lambda$4383, or
MgII~$\lambda$4481, are detected. The nebular continuum does not significantly
contribute to the optical emission, which is instead dominated by stars, indicating
that most massive stars have already exploded as supernovae.

$\bullet$ {\em Knot B} is a large cluster in which probably several knots are summed
together.  Although its spatial area comprises the Knot~3 in
emission lines, both peaks are displaced about 130~pc. The spectrum  increases
toward bluer wavelengths and shows marked absorption wings around the Balmer
lines in emission.

\begin{figure*}
\centering
\includegraphics[angle=0, width=1\linewidth]{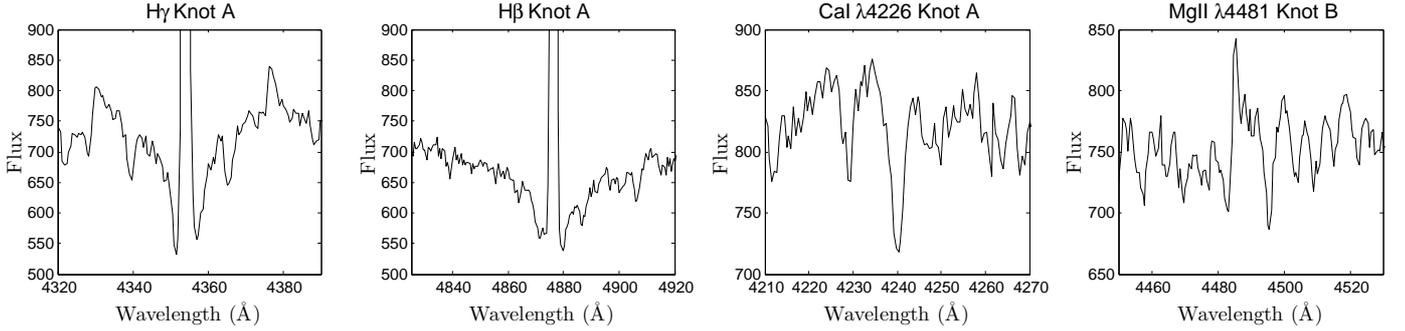}
\caption{Some of the absorption features detected in the knot~A: H$\gamma$ and
H$\beta$ absorption wings around H$\gamma$ and H$\beta$ in emission, 
CaI~$\lambda$4226 and 
MgII~$\lambda$4481. }
\label{Figure:Haro14-absor} 
\end{figure*}

Figure~\ref{Figure:Haro14spec}, covering a wide spectral range, does not allow us to
resolve individual features. As an example of the absorption features,
Figure~\ref{Figure:Haro14-absor} presents an enlarged version of particular regions of
the knot~A, where absorption lines are evident.

\subsubsection{Emission line fluxes}

\label{Section:fluxes}

We have measured the emission-line fluxes in each of the individual spectra using the {\sc
iraf}\footnote{IRAF= Image Reduction and Analysis Facility, is a software system for the reduction and
analysis of astronomical data. It is distributed by the NOAO, which is operated by the Association of
Universities for Research in Astronomy, Inc., under cooperative agreement with the National Science
Foundation} task {\sc splot}.  A Gaussian plus a linear function were fitted to the line and continuum
between two cursor-marked positions. In H$\gamma$ and H$\beta$ an underlying absorption is often
evident as wide absorption wings, in those cases we fitted two Gaussians, one in absorption and one in
emission, where the absorption wings were conspicuous. The equivalent width of H$\gamma$ and
H$\beta$ in absorption range between 2.5 and 5.0 \AA\  (see  Table~ \ref{tab:fluxes}), in good
agreement with the predictions of evolutionary synthesis models \citep{Olofsson1995,
GonzalezDelgado1999b}. 

Absorption wings are not visible around H$\alpha$; we then fixed the equivalent
width of H$\alpha$ in absorption equal to the value found for H$\beta$, which is a reasonable assumption following the predictions of the models \citep{Olofsson1995}.

\subsubsection{Reddening correction}

\label{Section:reddening}

We derived the interstellar reddening coefficient making use of the Balmer decrement, i.e.,
fitting the observed Balmer decrement to its theoretical value. Since the
Balmer line intensity ratios are precisely known from atomic theory, any deviation from them must
be due to the selective absorption by interstellar dust, which reduces the flux at shorter
wavelengths more efficiently than at longer wavelength. The extinction coefficient, C(H$\beta$),
is determined through the expression \begin{equation}
\frac{F_{\lambda}}{F(H_{\beta})}=\frac{F_{\lambda,0}}{F(H_{\beta,0})}\times10^{-C(H_{\beta})}\times(f(\lambda)-f(H_{\beta}))
,\end{equation} where F is the flux of any Balmer line in emission,
F$_{\lambda,0}$/F(H$_{\beta,0}$) is the ratio of fluxes that would be received on the Earth in
absence of interstellar extinction, F$_{\lambda}$/F(H$_{\beta}$)  is the ratio of the fluxes
actually observed, and f($\lambda$) is the extinction law.

We adopted the theoretical F(H$\alpha$)/F(H$\beta$)=2.87 value, assuming the case~B Balmer
recombination decrement in low-density limit, for a temperature of  10000~K \citep{Osterbrock2006}, and
we used the extinction law from \cite{ODonnell1994}.  This method assumes that the dust is in
a homogeneous obscuring screen, and that all the H$\alpha$ and H$\beta$ emission arises from hydrogen
recombination and not from collisional excitation.

Although H$\gamma$ falls into the observed spectral range, we only used the ratio H${\alpha}$/H${\beta}$
to derive the extinction coefficient because the H$\gamma$ flux is more severely affected by
underlying stellar absorption as compared to H$\alpha$ or H$\beta$; the strength of the Balmer
series lines in emission decreases rapidly with decreasing wavelength, whereas the equivalent width of
the stellar absorption is roughly constant \citep{Olofsson1995,GonzalezDelgado1999a}.

The derived interstellar extinction coefficient, C(H$\beta$), and the reddening-corrected intensity
ratios are presented in Table~\ref{tab:fluxes}. For an easier comparison with other data, the number
of magnitudes of extinction in V, A$_{V}$, and the color excess, $E(B-V)$,  are also shown.  It is
important to note that different regions show significantly different C(H$\beta$), from
0.034 in the Knot~{\sc b} to 0.735 in the Knot~{\sc a}; this translates into increments of magnitudes,
A$_{V}$, from 0.074 to 1.589.

\begin{landscape}
\begin{table}
\footnotesize
\caption{Reddening-corrected line intensity ratios.\label{tab:fluxes}}
\begin{tabular}{lcccccccccccc}
\hline
Ion               &\multicolumn{1}{c}{Knot 1}&\multicolumn{1}{c}{Knot 2}&\multicolumn{1}{c}{Knot 3}&\multicolumn{1}{c}{Knot 4}&\multicolumn{1}{c}{Knot 5}&\multicolumn{1}{c}{Knot 6}&\multicolumn{1}{c}{Knot 7}& \multicolumn{1}{c}{Knot 8} & \multicolumn{1}{c}{Knot A}&\multicolumn{1}{c}{Knot B} & \multicolumn{1}{c}{Integrated}\\
\hline\hline                                                                                                                                                                                                                                                                    
4340~H$\gamma$          &  0.492$\pm$0.009 &  0.473$\pm$0.007      &  0.509$\pm$0.014    &  0.498$\pm$0.013   &  0.537$\pm$0.024    &  0.538$\pm$0.017    &  0.435$\pm$0.006   &  ---                & 0.589$\pm$0.014  &   0.373$\pm$0.012    & 0.449$\pm$002      \\
4363~[OIII]             &  0.029$\pm$0.004 &  0.029$\pm$0.004      &    ---              &  ---               &   ---               &     ---             &  0.017$\pm$0.003   & ---                 &    ---           &   ---                & ---                \\
4861~H$\beta$           &  1.000  &  1.000   &  1.000   &  1.000  &  1.000 &  1.000   &  1.000  &   1.000   & 1.000 &    1.000  & 1.000    \\
4959~[OIII]             &  0.793$\pm$0.012 &  1.145$\pm$0.011      &  0.783$\pm$0.011    &  0.718$\pm$0.013   &  0.773$\pm$0.023    &  0.551$\pm$0.013    &  1.035$\pm$0.010   &  1.668$\pm$0.209    & 0.948$\pm$0.014  &   0.774$\pm$0.008    & 0.742$\pm$0.003    \\
5007~[OIII]             &  2.306$\pm$0.027 &  3.412$\pm$0.030      &  2.284$\pm$0.025    &  2.043$\pm$0.029   &  2.350$\pm$0.054    &  1.683$\pm$0.030    &  3.128$\pm$0.022   &   5.257$\pm$0.617   & 2.749$\pm$0.034  &  2.310$\pm$0.012     & 2.228$\pm$0.006    \\
6300~[OI]               &  0.079$\pm$0.005 &  0.032$\pm$0.003      &  0.031$\pm$0.003    &  0.076$\pm$0.005   &  0.093$\pm$0.011    &  0.076$\pm$0.007    &  0.040$\pm$0.003   &   0.328$\pm$0.064   & 0.045$\pm$0.005  &    0.071$\pm$0.010   & 0.100$\pm$0.001    \\
6312~[SIII]             &  0.013$\pm$0.004 &  0.014$\pm$0.003      &  0.012$\pm$0.003    &     ---            &   ---               &    ---              &  0.010$\pm$0.002   &  ---                &0.017$\pm$0.004   &        ---           & ---                \\
6364~[OI]               &  0.030$\pm$0.004 &  0.011$\pm$0.002      &  0.007$\pm$0.005    &     ---            &     ---             &  0.023$\pm$0.005    &     ---            &  ---                &    ---           &     ---              & ---                \\
6548~[NII]              &  0.090$\pm$0.005 &  0.064$\pm$0.003      &  0.074$\pm$0.004    &  0.141$\pm$0.006   &  0.108$\pm$0.009    &  0.110$\pm$0.006    &  0.065$\pm$0.003   &   ---               & 0.077$\pm$0.004  &    0.088$\pm$0.008   & 0.091$\pm$0.002    \\
6563~H$\alpha$          &  2.870$\pm$0.045 &  2.870$\pm$0.035      &  2.870$\pm$0.043    &  2.870$\pm$0.052   &  2.870$\pm$0.091    &  2.870$\pm$0.068    &  2.870$\pm$0.028   &   2.870$\pm$0.479   & 2.870$\pm$0.048  &    2.870$\pm$0.024   & 2.870$\pm$0.011    \\
6583~[NII]              &  0.278$\pm$0.008 &  0.192$\pm$0.004      &  0.215$\pm$0.006    &  0.422$\pm$0.011   &  0.323$\pm$0.014    &  0.326$\pm$0.012    &  0.193$\pm$0.005   &   0.275$\pm$0.060   & 0.240$\pm$0.006  &    0.292$\pm$0.009   & 0.300$\pm$0.002    \\
6678~HeI                &  0.030$\pm$0.004 &  0.032$\pm$0.003      &  0.033$\pm$0.004    &  0.029$\pm$0.005   &  0.015$\pm$0.008    &  0.024$\pm$0.005    &  0.030$\pm$0.003   &   ---               & 0.031$\pm$0.004  &      ---             & 0.024$\pm$0.001    \\
6716~[SII]              &  0.356$\pm$0.008 &  0.204$\pm$0.004      &  0.191$\pm$0.006    &  0.562$\pm$0.013   &  0.474$\pm$0.016    &  0.437$\pm$0.013    &  0.222$\pm$0.004   &    0.899$\pm$0.162  & 0.275$\pm$0.007  &   0.381$\pm$0.011    & 0.449$\pm$0.002    \\
6731~[SII]              &  0.254$\pm$0.007 &  0.145$\pm$0.004      &  0.138$\pm$0.005    &  0.402$\pm$0.010   &  0.329$\pm$0.014    &  0.319$\pm$0.011    &  0.167$\pm$0.004   &    0.646$\pm$0.121  & 0.196$\pm$0.006  &    0.281$\pm$0.010   & 0.321$\pm$0.002    \\
7065~HeI                &  0.013$\pm$0.003 &  0.021$\pm$0.003      &     ---             &      ---           &     ---             &   ---               &  0.021$\pm$0.003   &   ---               & 0.016$\pm$0.003  &       ---            & ---                \\
7136~[ArIII]            &  0.076$\pm$0.003 &  0.081$\pm$0.003      &  0.076$\pm$0.004    &  0.073$\pm$0.005   &  0.054$\pm$0.008    &  0.051$\pm$0.005    &  0.072$\pm$0.003   &  ---                & 0.075$\pm$0.004  &    0.070$\pm$0.007   & 0.060$\pm$0.001    \\
7330~[OII]              &  0.035$\pm$0.004 &  0.020$\pm$0.002      &  0.018$\pm$0.005    &   ---              &  0.024$\pm$0.009    &   ---               &  0.022$\pm$0.002   &  ---                & 0.019$\pm$0.005  &   ---                & ---                \\
\hline\hline
W(H$\gamma$)$_{ab}$     &  0                &  2.5                  &           3.0       &     0              &   4.0               & 3.8                 &       4.3          & ---                 & 3.6              &         4.4          & 5.7                \\
W(H$\beta$)$_{ab}$      &  2.8              &  3.6                  &           5.0       &     5.0            &   4.6               & 4.7                 &       2.8          & 2.6                 & 4.5              &        5.0           & 3.5                \\\hline
F$_{H\beta}$            &  71.7$\pm$2.4     &  149.3$\pm$3.7        &  191.4$\pm$15.3     &  78.7$\pm$4.4      &  88.0$\pm$12.5      &  85.3$\pm$6.7       &  102.0$\pm$2.4     & 10.61$\pm$0.9       &  317.6$\pm$38.4  &  45.8$\pm$0.7        & 2530.3$\pm$27.8  \\
W(H$\beta$)             &  45$\pm$1.0       &  30.9$\pm$0.3         &   18.0$\pm$0.2      &  23.3$\pm$0.3      &  5.1$\pm$0.1        & 14.3$\pm$0.3        &  47.5$\pm$0.4      & 1.9$\pm$0.1         &  7.0$\pm$0.1         &  6.2$\pm$0.1     & 7.4$\pm$0.1 \\
C$_{H\beta}$            &  0.268$\pm$0.013  &  0.246$\pm$0.010      &  0.618$\pm$0.012    &  0.411$\pm$0.015   &  0.552$\pm$0.026    &  0.440$\pm$0.020    &  0.167$\pm$0.008   & 0.621$\pm$0.139     &  0.735$\pm$0.014 &    0.034$\pm$0.007   & 0.379$\pm$0.003    \\
$A_{V}$                 &  0.579$\pm$0.028  &  0.531$\pm$0.022      &   1.336$\pm$0.027   &  0.888$\pm$0.033   &  1.193$\pm$0.057    &  0.950$\pm$0.043    & 0.360$\pm$0.017    & 1.342$\pm$0.300     & 1.589$\pm$0.030  &    0.074$\pm$0.015   & 0.819$\pm$0.007    \\                                                                                                                                                                                                   
E(B-V)                  &  0.187$\pm$0.009 &  0.171$\pm$0.007      &   0.431$\pm$0.009   &  0.286$\pm$0.011   &  0.385$\pm$0.018    &  0.307$\pm$0.014    &  0.116$\pm$0.006   & 0.433$\pm$0.097     & 0.512$\pm$0.010  &    0.024$\pm$0.005   & 0.264$\pm$0.002    \\
\hline
\end{tabular}
\end{table}
Notes.- Reddening-corrected line fluxes normalized to F(H$\beta$)=1 for the individual regions selected in Haro\,14.
The values of the equivalent width in absorption for H$\gamma$ and H$\beta$ 
are provided in the second part of the table.
The reddening-corrected H$\beta$
flux (in units of 10$^{-16}$erg~s$^{-1}$~cm$^{-2}$), the H$\beta$ equivalent width in emission,
 the
interstellar extinction coefficient,C$_{H\beta}$, the color excess, E(B-V), and A$_{V}$
for each region are also tabulated. A$_{V}$ was derived
using  A$_{V}$=2.16075$\times$C(H$\beta$).      
\end{landscape}

\begin{landscape}
\begin{table}
\footnotesize
\caption{Line ratios, physical parameters, and abundances.\label{tab:diagnostic}}
\begin{tabular}{lcccccccccccc}
\hline
Parameter                                               &\multicolumn{1}{c}{Knot 1}       &\multicolumn{1}{c}{Knot 2}      &\multicolumn{1}{c}{Knot 3}        &\multicolumn{1}{c}{Knot 4}     &\multicolumn{1}{c}{Knot 5}       &\multicolumn{1}{c}{Knot 6}         &\multicolumn{1}{c}{Knot 7}        & \multicolumn{1}{c}{Knot~8}    &\multicolumn{1}{c}{Knot A}            &\multicolumn{1}{c}{Knot B} & \multicolumn{1}{c}{Sum}\\
\hline
$[\ion{O}{iii}]~\lambda5007/\Hb$                         &    2.305$\pm$0.024        &  3.412$\pm$0.027          &   2.285$\pm$0.022         &   2.043$\pm$0.027         & 2.349$\pm$0.047           &  1.683$\pm$0.026            & 3.128$\pm$0.020              &  5.258$\pm$0.617          &  2.749$\pm$0.030                &   2.310$\pm$0.012    & 2.223$\pm$0.006 \\
$[\ion{O}{i}]~\lambda6300/\Ha$                           &    0.027$\pm$0.002        &  0.011$\pm$0.001          &   0.011$\pm$0.001         &   0.027$\pm$0.002         & 0.032$\pm$0.004           &  0.026$\pm$0.002            & 0.014$\pm$0.001              &  0.114$\pm$0.029          &  0.016$\pm$0.002                &   0.025$\pm$0.003    & 0.035$\pm$0.001 \\
$[\ion{N}{ii}]~\lambda6584/\Ha$                          &    0.097$\pm$0.002        &  0.067$\pm$0.001          &   0.075$\pm$0.002         &   0.147$\pm$0.003         & 0.112$\pm$0.003           &  0.114$\pm$0.003            &  0.067$\pm$0.001             &  0.096$\pm$0.026          &  0.083$\pm$0.002                &   0.102$\pm$0.003    & 0.104$\pm$0.001 \\
$[\ion{S}{ii}]~\lambda\lambda6717\,6731/\Ha$             &    0.212$\pm$0.003        &  0.122$\pm$0.002          &   0.115$\pm$0.002         &   0.336$\pm$0.004         & 0.279$\pm$0.005           &  0.263$\pm$0.004            &  0.135$\pm$0.002             &  0.538$\pm$0.114          &  0.164$\pm$0.003                &   0.231$\pm$0.005    & 0.269$\pm$0.001  \\\hline 
N$_{e}$   (cm$^{-3}$)                                    &   $<$100                   &    $<$100                &  $<$100                    &  $<$100                  &     $<$100                &    $<$100                 &    $\approx$ 100             &     $<$100               & $<$100                         &       $<$100           &   $<$100 \\
T$_{e}$  (K)                                             &    12500$\pm$220          &  10900$\pm$190           &    ---                     &      ---                 &       ---                &         ---              &    9500$\pm$200    &         ---                  &     ---                       &   ---                    & ---    \\
12+log(O/H)$^{2}$                                        &    8.21                  &  8.22                    &  8.26                      &   8.24                 & 8.21                       &  8.22                       &  8.20                    &    ---                        &  8.22                 &   8.21           &  8.19\\
\hline\hline 
\end{tabular}
Notes.-  Abundances are derived following 
 \cite{PilyuginGrebel2016}.
\end{table}
\end{landscape}

\normalsize

\subsubsection{Diagnostic line ratios, physical parameters, and oxygen abundances}
\label{Section:diagnostic}

\begin{figure*} 
\centering 
\includegraphics[angle=0,width=0.8\linewidth]{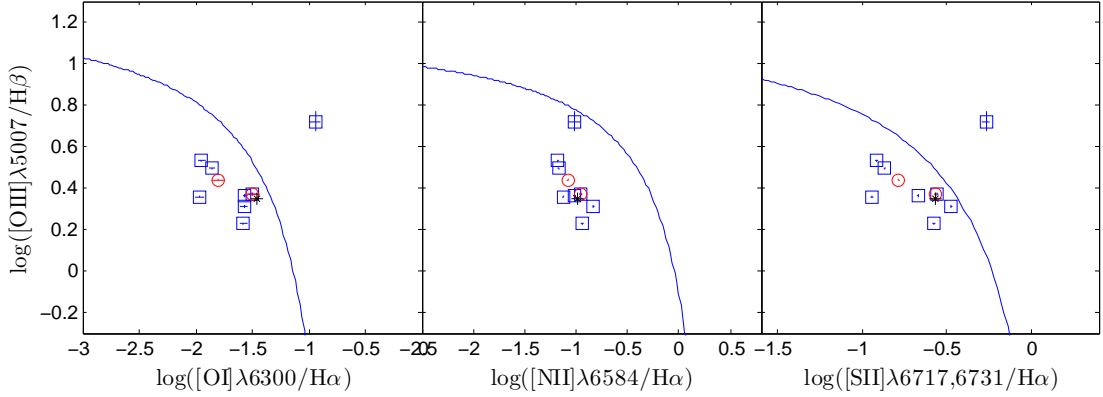} \caption{
Optical emission-line diagnostic
diagram for the individual regions identified in Haro\,14, with the theoretical 
maximum starburst line \citep{Kewley2001}. 
Ratio values for the eight SF knots 
are shown as blue squares, for the two HSB stellar clusters as red circles, and 
for the integrated spectrum as a black star; errors are plotted as small crosses.}
\label{Figure:diagnostic} 
\end{figure*}

Values of the diagnostic ratios for the regions selected in Haro\,14 are shown in
Table~\ref{tab:diagnostic}. Figure~\ref{Figure:diagnostic} shows these values in the diagnostic
diagrams [\ion{O}{iii}]~$\lambda5007$/\Hb{} {\em versus} [\ion{N}{ii}]~$\lambda6584$/\Ha,
[\ion{S}{ii}]~$\lambda\lambda6717,\,6731$/\Ha, and [\ion{O}{i}]~$\lambda6300$/\Ha\ . As expected, all
the major SF regions fall in the zone of the diagram corresponding to photoionization by hot stars. But
Knot~8, the faint blob detected at the galaxy west, lies well above the line delimiting the starburst
area on the [\ion{O}{iii}]~$\lambda5007$/\Hb{} versus [\ion{O}{i}]~$\lambda6300$/\Ha\, and
[\ion{O}{iii}]~$\lambda5007$/\Hb{} versus [\ion{S}{ii}]~$\lambda\lambda6717,\,6731$/\Ha\ diagrams.

The electronic density and temperature and oxygen
abundances of the ionized gas in the individual regions were derived from the reddening-corrected emission-line fluxes.

Electron densities were computed using the
[\ion{S}{ii}]~$\lambda\lambda6717,\,6731$ ratio  \citep{Osterbrock2006}; the derived
values are shown in Table~\ref{tab:diagnostic}. The densities in all regions fall in
the low-density limit regime ($\leq$100cm$^{-3}$), except in Knot~7, where it is
slightly higher ($\approx$ 100cm$^{-3}$).

The electron temperature was calculated from the
[\ion{O}{iii}]$\lambda$4363/($\lambda$4959+$\lambda$5007) ratio in those cases in which reliable values
of [\ion{O}{iii}]~$\lambda$4363 were measured, i.e., Knots~1,~2, and ~7. We follow the procedure  described
in \cite{IzotovStasinska2006}. The derived values are shown in Table~\ref{tab:diagnostic}.  In
Knot~7 where the electronic density is slightly higher, the temperature found is correspondingly
lower.

The oxygen abundance cannot be derived using the most reliable direct-T$_{e}$ method, not even  
in the knots where the [\ion{O}{iii}]$\lambda$4363 has been
measured.  The direct-T$_{e}$ method requires the fluxes of the
[\ion{O}{ii}]~$\lambda$3727+3729 lines or, alternatively, the
[\ion{O}{ii}]~$\lambda$7320,7331 lines. The VIMOS observing spectral range does not
include the most commonly used [\ion{O}{ii}]~$\lambda$3727+3729 lines. The
[\ion{O}{ii}]~$\lambda$7320,7331 lines, although within the observed range, fall in
the zone of the spectra that is badly affected by sky residuals, which makes it impossible to
obtain reliable values of the [\ion{O}{ii}]~$\lambda$7320 flux. 

Numerous alternative methods were suggested for deriving abundances in those cases for which
the direct Te-method is not applicable. We estimated the oxygen abundance using the
expressions recently proposed by \cite{PilyuginGrebel2016}. These authors provide a calibration
that is particularly indicated for cases for which the lines [\ion{O}{ii}]~$\lambda$3727+3729 lines are
not available: the oxygen abundance is derived from the intensities of the strong lines
[\ion{O}{iii}]$\lambda$$\lambda$4957,5007, [\ion{N}{ii}]$\lambda$$\lambda$6548,6584, and
[\ion{S}{ii}]$\lambda$$\lambda$6717,6731. 

Results for the regions selected in Haro\,14 are shown in Table~\ref{tab:diagnostic}. No
signficant spatial variations in the abundances were found; the relative accuracy of the abundances based on the \cite{PilyuginGrebel2016} 
calibration is 0.1~dex. The derived
value for the oxygen abundance, 12+log(O/H)$\approx$8.22$\pm$0.1, is slightly lower than the
value of 12+log(O/H)$\approx$8.41$\pm$0.02 reported in \cite{Hunter1999}.

\subsection{Kinematics of the ionized gas}

We studied the kinematics of the ionized gas fitting Gaussian curves to the line profiles of
the brightest emission lines, namely, H$\alpha$ and [\ion{O}{iii}]$\lambda$5007. A
significant percentage of the high S/N H$\alpha$ line profiles exhibit a broad low-intensity
component and, in those cases, the profiles were optimally fitted using two Gaussians: a
bright narrow component (with $\sigma$$\leq$ 35~km~s$^{-1}$, corrected for instrumental
broadening), and the broad low-intensity component (80~km~s$^{-1}$ $\leq$ $\sigma$ $\leq$
150~km~s$^{-1}$, also corrected for instrumental effects). The broad component is quite
faint and hard to fit even in H$\alpha$.

\begin{figure*}
\centering
\includegraphics[angle=0, width=0.8\linewidth]{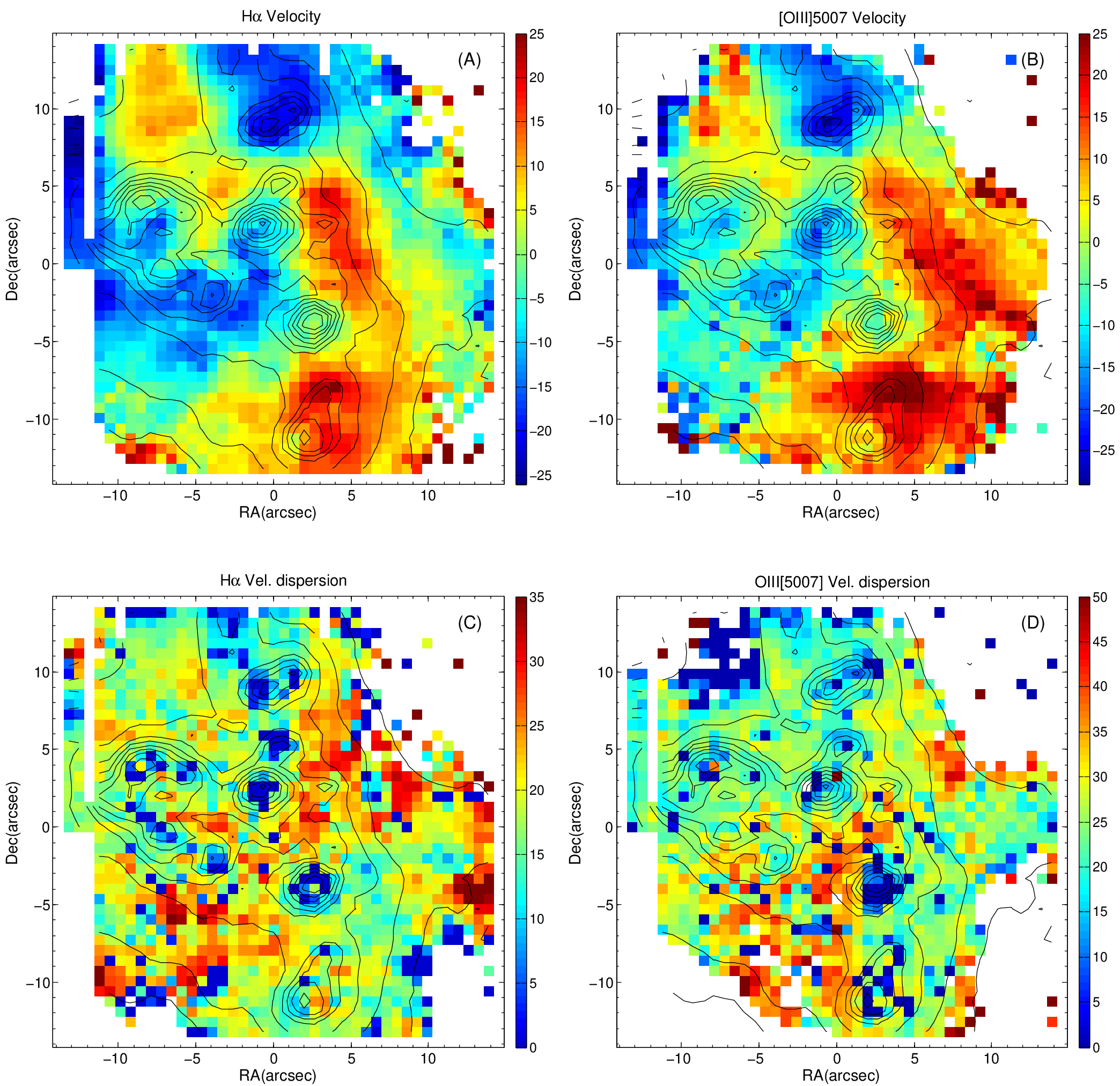}
\caption{H$\alpha$ and [\ion{O}{iii}]$\lambda$5007 line-of-sight velocity
fields (panels A and B) and velocity dispersion maps (panels C and D).}
\label{Figure:velocitymaps} 
\end{figure*}

The line-of-sight velocity of the emitting gas in H$\alpha$ (for the bright narrow component) and
[\ion{O}{iii}]$\lambda$5007, measured from the Doppler shift of the Gaussian line profile centroids
relative to the systemic velocity of the galaxy, are shown in Figure~\ref{Figure:velocitymaps}. In
both maps the velocity distribution presents a complex structure: most regions are moving with
respect to the systemic velocity, but the movements do not correspond to a simple rotation. Regions
of material are moving toward us in the east and north galaxy areas: at the north, the blueshifted
region spatially coincides with the Knot~1 and, at the east, the region traces the periphery of the
horseshoe-like curvilinear structure. Such complex velocity fields trace the movements of the gas in
the perturbed ISM that is dominated by feedback effects.  The amplitude of the velocity field is about
50~km~s$^{-1}$. H$\alpha$ and [\ion{O}{iii}]$\lambda$5007 velocity fields display roughly the same
pattern, but discrepancies appear on the galaxy west, where some redshifted regions come out in
[\ion{O}{iii}]$\lambda$5007 but not in  H$\alpha$. Differences between the hydrogen recombination and
forbidden line velocities maps can arise because these lines did not originate in the same gas
region, and Knot~8 is better distinguished in [\ion{O}{iii}]$\lambda$5007.  

Figure~\ref{Figure:velocitymaps} also shows the line-of-sight velocity dispersion for the bright
narrow component of H$\alpha$ and [\ion{O}{iii}]$\lambda$5007. Values of the dispersions are low in
the whole covered area, but the map has a clear structure: the regions with higher dispersions (
25~km~s$^{-1}$ $\leq$ $\sigma$ $\leq$ 35~km~s$^{-1}$) are situated in the periphery of the major SF
knots, and  are mostly concentrated around a band, which crosses the galaxy in the southeast
to northwest direction. Also, shocks dominated areas have higher line-of-sight velocity dispersions
and, as expected, \ion{H}{ii} regions present the lowest dispersion values.

\begin{figure*}
\centering
\includegraphics[angle=0, width=1.05\linewidth]{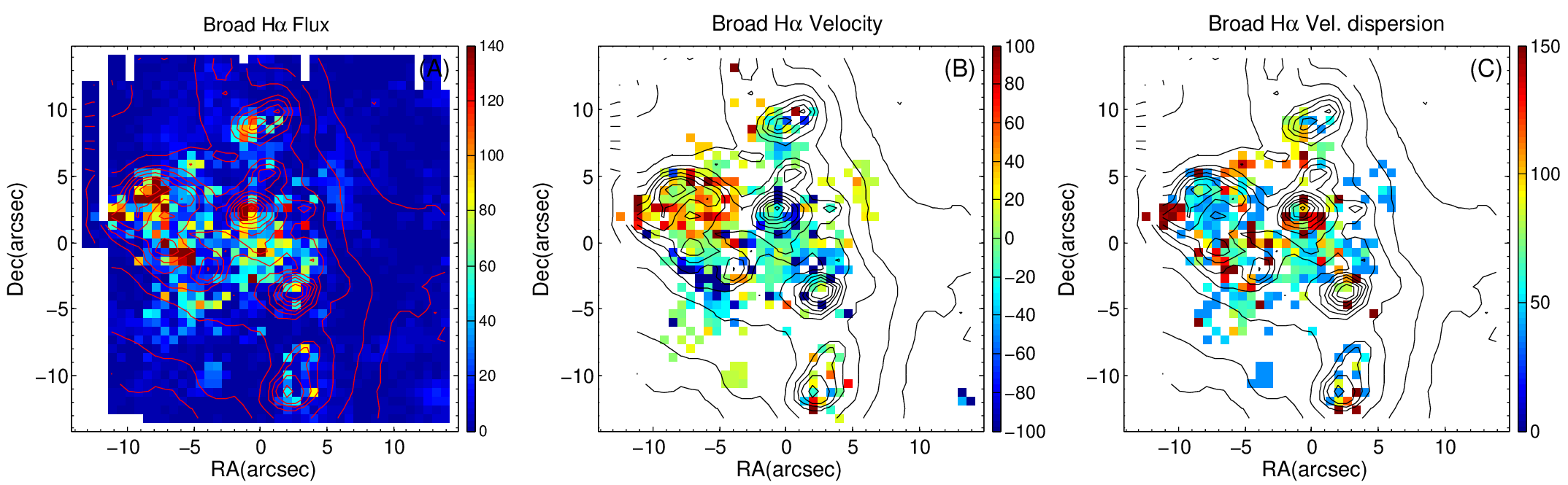}
\caption{H$\alpha$ flux for the low-brightness broad component (panel A). H$\alpha$ velocity field
 for the low-brightness broad component (panel B).
Velocity dispersion for the low-brightness broad component (panel C).}
\label{Figure:map-wide-compo} 
\end{figure*}

The H$\alpha$ emission flux map of the broad low-intensity component is shown in 
Figure~\ref{Figure:map-wide-compo}. This second component is faint (its luminosity represents
about 6\% of the total H$\alpha$ luminosity), but the structure of its flux distribution
confirms that it is indeed a real component and not the result of instrumental effects. The
emission is localized in the major regions of SF, being stronger in the knots placed in the
periphery of the superbubble feature. The line-of-sight velocity and velocity dispersion for
the broad low-intensity component are also shown in Figure~\ref{Figure:map-wide-compo}.

Multiple components have often been found in BCGs 
\citep{Ostlin2001,Westmoquette2008,Westmoquette2010,James2009,James2013} and, in particular, the
presence and origin of a low-intensity broad spectral component has been widely discussed.
\cite{Izotov2007b} conclude, from an analysis of a large number of BCGs exhibiting broad emission
lines, that most probably this emission is associated with the evolution of massive stars and
their interaction with the circumstellar and ISM. The spatial distribution of the 
H$\alpha$ flux (Panel (A) in Figure~\ref{Figure:map-wide-compo}) is consistent  with this
scenario, as the emission is restricted to the major SF regions and their close proximity.   The
values of the line-of-sight velocity dispersions (80~km~s$^{-1}$ $\leq$ $\sigma$ $\leq$
150~km~s$^{-1}$) are also consistent with this broadband feature caused by dynamical processes
related to the evolution of massive stars.

\section{Discussion}

The first step toward establishing the evolutionary status and star-forming history of a galaxy
is to disentangle and characterize its different stellar components.  The comparison of
continuum and emission-line maps provide us with the first insights into the stellar content
of Haro\,14. Three different stellar populations were distinguished: the very
young stars (i.e., Knots~1-7), an intermediate-age stellar population (Knots~{\sc a}
and {\sc b}), and the more extended, redder, and supposedly older underlying stellar
component. The small and fainter blobs detected at large galactocentric distances 
(i.e.,
Knot~8, see Figure~\ref{Figure:Ha-contour}) could be even younger sites of SF. 

The stars detected in emission-line maps are young, as only short-lived OB stars,  with
temperatures $\geq$ 30.000~K and masses above 10 M$\odot$, are able to produce photons with
energy large enough to ionize hydrogen ($\geq$13.6~eV); because such stars evolve quickly, on
timescales of Myrs, emission-line maps only trace stars with ages $\leq$ 10 Myrs.

Haro\,14 emission-line maps revealed  SF through the whole mapped area. Using the H$\alpha$
map we identified and delimited eight major knots: four \ion{H}{ii}-regions located in
the chain-like structure oriented north-south, and another three situated in the rim of the
loop, which departs from the central chain and stretches toward east (see
Section~\ref{Section:morphology}).  At LSB levels several blobs and
filaments were barely distinguished;  the major blob is located at the galaxy west (Knot~8).

The morphology in continuum and emission lines (Figure~\ref{Figure:HaroCont}) strongly advocate a scenario of triggered SF, with the star-forming
activity propagating outward. The emission in continuum peaks close to the galaxy center,
whereas the maxima in emission lines are reached at larger galactocentric distances. According
to this scenario the recent episode of SF would have been triggered by the mechanical energy
input from massive stars and supernovae generated in a previous burst (likely the HSB continuum
clusters); the knots situated in the loop (namely Knots~5, 6 and 7) strongly suggest new SF that is taking place in the walls of a large-scale expanding structure (supershell or
superbubble), and the fainter knots at larger galactocentric distances could be even younger
sites of SF.

Triggered SF occurs when supersonic flows generated by distant supernovae blast waves or
stellar winds sweep up a shell of shocked neutral gas; eventually this shell fragments and
condenses to produce a new generation of stars \citep{Elmegreen1977}. If triggered SF has happened, the separation
in time and space of the various episodes of SF must be related in a consistent way. Therefore
the first step to verify the scenario of triggered SF in Haro\,14 is to constraint the ages of
its different episodes of SF, namely, the young \ion{H}{ii}-regions (Knots~1-8) and the HSB
stellar clusters.

Table~\ref{SF-KnotI} presents the H$\alpha$ sizes, fluxes, equivalent widths, and luminosities of the
SF knots; their sizes go from 131 up to 231~pc, placing them in the range of giant \ion{H}{ii}
regions. As a reference, 30~Doradus has a diameter of about 300~pc; these diameters must be
understood only as crude indicators of size. We are limited in resolution and some knots, such as Knot~2,
include most probably at least two different SF regions. Indeed, it is not easy to  distinguish
between large simple \ion{H}{ii} regions and close clusters of smaller regions, which is beyond the
scope of this work.   All but Knot~8 show H$\alpha$ luminosities in the range of giant extragalactic
\ion{H}{ii} regions. Star-forming rates (SFR) and number of ionizing photons, computed from the
H$\alpha$ luminosity following \cite{Osterbrock2006}, are also shown.

We used the {\sc Starburst~99} evolutionary synthesis models \citep{Leitherer1999} to
constraint the properties of the knots.  We compared the measured H$\alpha$ equivalent widths
with the values predicted by models with metallicity z=0.008, which is the value closer to the
metallicity derived from the emission-line fluxes, a reasonable approximation to the
metallicity of a young population. We found that, in all cases, we can reproduce the
equivalent widths with an instantaneous burst of SF and the Salpeter initial mass
function (IMF) with an upper mass limit of 100 M$\odot$. The ages estimated for each knot are 
presented in Table~\ref{SF-KnotI}.

These age estimates are upper limits to the real ages, as the measured equivalent
widths of Balmer emission lines are always inferior levels to the true values. Two
factors can severely affect the measurement of the equivalent width in emission: first, the
contribution of the underlying absorption due to AF stars from the burst, which is
included in the emission flux and decreases its final value; and, second, the presence of a
higher continuum due to older stars, which dilutes the values of
the equivalent width. The decrease of equivalent width in turn results in larger
ages.

Measured H$\alpha$ fluxes were corrected for stellar absorption (see
Section~\ref{Section:fluxes}), but no attempt has been carried out to correct for the  effect
of a higher continuum; such a correction would require a precise modeling of the older and
intermediate-age stellar populations in every spaxel.

Most of the regions present ages around 6~Myr, which is consistent with the
non-detection of Wolf-Rayet signatures in the integrated spectra of the knots. Only
Knot~5 has noticeably larger ages, but the spectrum of this knot shows a substantially high continuum and, at the same
time, it is moderate in flux (the larger the flux in emission, the less important the
effects of absorption and dilution). Therefore, the larger ages found here are most
probably due to the effect of the underlying population.

\begin{table*}
\caption{H$\alpha$ properties of the SF knots}
\begin{center}
\begin{tabular}{|c|c|c|c|c|c|c|c|}
\hline
Knot   & Diameter                 & F(H$_{\alpha}$)             &  W(H$_{\alpha}$)  & logL(H$_{\alpha}$)  & SFR                                      & N$_{LyC}$                                  &   Age$_{H_{\alpha}}$   \\
       & (pc)                     & (erg cm$^{-2}$s$^{-1}$)     & (\AA)      &  (erg s$^{-1}$)     & $\times$10$^{-3}$(M$\odot$yr$^{-1}$)     & $\times$10$^{50}$ (photons  s$^{-1}$)                     &       Myrs               \\\hline\hline
1      &  190                     & 205$\pm$4                   & 177              & 38.62               & 3.2                                    & 3.02                                           & 5.7                     \\
2      & 144                      & 428$\pm$6                   & 157              & 38.94               & 6.7                                    & 6.24                                          & 5.9          \\
3      & 131                      & 550$\pm$17                  & 111              & 39.05               & 8.5                                    & 7.85                                          & 6.2       \\
4      & 186                      & 226$\pm$6                   & 150              & 38.67               & 3.5                                    & 3.26                                          &  5.9         \\
5      & 169                      & 253$\pm$13                  & 28               & 38.71               & 3.5                                    & 3.24                                          &  8.5               \\
6      & 152                      & 245$\pm$9                   & 75               & 38.69               & 3.71                                   & 3.44                                          &  6.6          \\
7      & 158                      & 292$\pm$3                   & 223              & 38.77               & 4.64                                   & 4.29                                                &  5.4            \\
8      & 231                      & 30$\pm$9                    & 11               & 37.78               & 0.4                                    & ---                                                 & ---            \\
\hline
\end{tabular}
\end{center}
Notes: F(H$_{\alpha}$)
in units of 10$^{-16}$erg~s$^{-1}$~cm$^{-2}$. H$\alpha$ fluxes
were corrected for interstellar extinction using the values provided in
Table~\ref{tab:fluxes}. 
\label{SF-KnotI}
\end{table*}

Knots~{\sc a} and {\sc b}  (Figure~\ref{Figure:HaroCont}) do not spatially coincide with any
peak in emission lines, indicating that 
the SF activity has already ceased.  Therefore no reliable ages can be derived using equivalent width of Balmer lines in emission, as
most of the light is coming from the nearby \ion{H}{ii} regions.  Alternatively, equivalent widths
in absorption of the higher order Balmer lines have been shown to be reliable age indicators
in post-starburst galaxies \citep{Olofsson1995, GonzalezDelgado1999b}. By comparing the equivalent
width in absorption of H$\beta$ and H$\gamma$ with the predictions of the evolutionary
synthesis models by \cite{GonzalezDelgado1999b}, we found ages between 6 and 15~Myrs  (for an
instantaneous burst of solar metallicity) and between 15 and 30 Myrs (for an instantaneous
burst of 0.05~Z$\odot$) for Knots~{\sc a} and {\sc b}.  The NIR colors of
these knots (J-H=0.6 and H-K$_{s}$=0.2; \citealp{Noeske2003}) appear as earliest
when the NIR becomes dominated by red supergiants, indicating ages between 10 and 30 Myrs, in
good agreement with our findings.

We find ages younger than 6~Myr for the extended starburst in Haro\,14,  whereas the
central HSB clusters are fairly older --- ages of 10-30~Myrs. Adopting an age of 20~Myrs
for the HSB clusters, we find that a blast wave originated in supernovae explosions must travel with
a velocity $v$= 40~km~s$^{-1}$ to ignite a SF event at $\approx$ 590~pc (the position of
Knot~1)  5.7~Myrs ago. To ignite the SF at the position of Knot~2 (at $\approx$ 168~pc), a
velocity of about 12~km~s$^{-1}$ is required. Such velocities are in good agreement with the
range of velocities found for expanding shells in dwarf irregular and star-forming galaxies
\citep{Walter1999,Silich2006,Egorov2014}. These calculations  are just a  
very plain approximation. The real case is obviously much more complex, as many supernovae explosions
are taking place in different spatial positions and at different times, and the final
superbubbles are the collective effect of all of these explosions.

Knot~8 and the other small condensations detected at the southeast periphery of the
field, are also suggestive of a scenario of triggered SF. Their high-excitation values are
consistent with very young ages; stellar evolution models predict a drop in
[\ion{O}{iii}]$\lambda$5007/H$\beta$ for ages $\geq$4~Myrs \citep{Copetti1986,Stasinska1996},
as the most massive stars leave the main sequence. And also shocks are expected to increase
[\ion{O}{iii}]$\lambda$5007 \citep{Raymond1988}.  On the other hand, a blast wave originated
in the central HSB clusters and traveling with a velocity about 40~km~s$^{-1}$ takes
about 18~Myrs to reach Knot~8.

A substantial percentage of the galaxy spaxels, mostly the regions situated at the southeast
and west, are being ionized by a mechanism other than hot stars (see
Figure~\ref{Figure:diagnostic-spaxel}). This is also in agreement with a scenario of
triggered SF because in star-forming regions  the most viable mechanisms to drive
non-photoionization processes are shocks from supernovae and massive stars winds \citep{Shull1979,
Allen1999}.

Summarizing,  we conclude that in Haro\,14 we most probably see a feedback mechanism at
work, i.e., we are able to distinguish different episodes of 
sequentially triggered (star-induced) SF driven by mechanical energy input from
concentrations of massive stars.

\section{Conclusions}

We present a comprehensive spectrophotometric study of the BCG Haro\,14 based on IFS
data taken with VIMOS at the VLT. About the central 1.7 $\times$ 1.7~kpc$^{2}$  of Haro\,14 
was observed in the 4150\dots7400\,\AA\ wavelength range with a spatial sampling of
$0\farcs66$ per spaxel. 

From these data, we build maps in continuum and in the brighter emission lines (e.g., H$\beta$,
[\ion{O}{iii}]~$\lambda5007$, H$\alpha$, [\ion{N}{ii}]~$\lambda6584,$ or
[\ion{S}{ii}]~$\lambda\lambda6717,\,6731$), which provide information about the galaxy
morphology and give insights into its stellar content; produce the H$\alpha$/H$\beta$ Balmer
decrement map, which depicts the dust;   build the
[\ion{S}{ii}]~$\lambda6717$/[\ion{S}{ii}]~$\lambda6731$ line-ratio map, which traces the 
electronic density; construct diagnostic line-ratio maps, namely,  
[\ion{O}{iii}]~$\lambda5007$/\Hb{}, [\ion{N}{ii}]~$\lambda6584$/\Ha,
[\ion{S}{ii}]~$\lambda\lambda6717,\,6731$/\Ha,{} and  [\ion{O}{i}]~$\lambda6300$/\Ha{}, which
are used together with diagnostic diagrams to carry out an extensive investigation of the 
nebula ionization and excitation; generate the integrated spectrum of the major \ion{H}{ii}
regions and young stellar clusters identified in the maps to determine reliable physical
parameters and oxygen abundances; and, finally, build the velocity and velocity dispersion
fields, which trace the ionized gas kinematics. The analysis of these datasets allows us to
derive the following conclusions:

$\bullet$ Haro\,14 presents significant deviations between its stellar and ionized gas distribution.
In continuum the intensity peaks close to the galaxy center, where two major stellar clusters have
been identified --- Knots~{\sc a} and {\sc b}. In emission line, maxima are distributed in the whole
mapped area, where the major SF regions are placed along a linear (chain-like) structure, elongated in
the north-south direction, and in a horseshoe-like curvilinear feature, which departs from the
central region and extends about $12\farcs0$ (760 pc) eastward; several fainter lumps, better 
distinguished
in the high-ionization [\ion{O}{iii}]$\lambda$5007 map, 
come out around the major \ion{H}{ii} regions, at distances about 0.7-0.8~kpc from the galaxy center;
the largest (Knot~8) is located at the west.   

Using the H$\alpha$ maps we delimited eight major SF knots, none of which is cospatial with the
continuum peaks. This morphology is suggestive of a mechanism of triggered SF taking place in
Haro\,14. 

$\bullet$ We disentangle two different episodes of SF in the central regions of Haro\,14. By
comparing with evolutionary synthesis models, we constrain their ages:  the recent starburst
(Knots~1-7) present ages $\leq$6~Myrs, whereas the HSB stellar clusters (Knots~{\sc a} and {\sc b})
are appreciably older, with ages between 10 and 30~Myrs. These stellar components rest on a
several Gyr old underlying LSB host galaxy \citep{Marlowe1999,Noeske2003}.

$\bullet$ The galaxy presents a highly inhomogeneous H$\alpha$/H$\beta$ line-ratio pattern,
in which the SF regions are practically dust free, whereas very large H$\alpha$/H$\beta$ values
appear in the periphery of the knots; the color excess varies from $E(B-V)$=0.04 up to
$E(B-V)$=1.09, when moving from the SF center to the galaxy outskirts. Such large variations
in $E(B-V)$ point out the importance of obtaining two-dimensional information on the dust
distribution: the use of an unique extinction coefficient for the whole galaxy (as usually
done when the spectroscopic data are taken using a slit) can yield large errors in the
extinction correction and, hence, in the derived fluxes and magnitudes.

$\bullet$ Diagnostic maps and diagnostic diagrams reveal the presence of a mechanism other than
photoionization by hot stars in a significant area of the galaxy. Shocks are the dominant ionizing
mechanism in the galaxy southeast and western regions, and also shock-ionized ring and arc features
are found surrounding the major SF knots.

$\bullet$ The excitation map reveal an interesting pattern; the SF knots show, as
expected, high-excitation values, but there are also three other zones with very high
excitation ([\ion{O}{iii}]~$\lambda5007$/\Hb{} up to six), which do not spatially
coincide with any of the major knots. Two of these high-excitation regions are
associated with faint blobs seen in the galaxy outskirts, and the third is an
arc feature located at the south of the chain-like structure, between Knots~3 and 4.
These three regions are dominated by shocks.  Also very young ages can contribute to
such high excitations.

$\bullet$ We produce the integrated spectrum of the eight major SF regions and the two HSB
stellar clusters; using these spectra we derive reliable physical properties and oxygen
abundances of the 10 selected knots. Electronic densities are found to be low
(N$_{e}\leq$100cm$^{-3}$) in all the regions but in Knot~7, which presents values slightly
higher (N$_{e}\approx$100cm$^{-3}$). We derived an oxygen abundance of 
12+log(O/H)$\approx$8.22$\pm$0.1 and found no evidence for any significant chemical abundance
gradient.

$\bullet$ The kinematics of the warm ionized gas was studied by fitting Gaussian curves to the
line profiles of the brightest emission lines, namely, H$\alpha$ and [\ion{O}{iii}]$\lambda$5007. The
velocity distribution presents a complex structure with regions of material moving toward us in the
east and north galaxy areas. Such complex velocity fields trace the movements of the gas in a
perturbed ISM, dominated by feedback effects. Ages derived for the current SF episode (around 6~Myr)
are consistent with   the energetics of the ISM dominated by core-collapse Type-II~supenovae explosions. The
amplitude of the velocity field is about 50~km~s$^{-1}$.

In conclusion,  the information derived from the VIMOS data allows us to probe the recent
star-forming history of Haro\,14. We disentangle two temporally and spatially  separated star-forming
episodes: the recent starburst (Knots~1-7) and the HSB stellar clusters (Knots~{\sc a} and {\sc b}).
The morphological pattern, the irregular velocity field, dominated by the movement of the ionized
gas, and the presence of shocks, all point to a scenario of triggered (star-induced) star
formation. The ages of the different episodes of SF, $\approx$ 6~Myr for the extended starburst in
Haro\,14  and about 10-30~Myrs for the central HSB clusters, are consistent with the ongoing
burst being triggered by the collective action of stellar winds and supernovae that originated in the central HSB
stellar clusters.

\begin{acknowledgements} LMC acknowledges support from the Deutsche Forschungsgemeinschaft
(CA~1243/1-1). The authors are very grateful to N. Caon and P. Weilbacher, members of the IFU-BCG project.
We also thank R. Manso-Sainz for many constructive discussions and comments, and R. Mettin for a careful reading
of the manuscript. Further thanks goes to Oliver Mettin for extremely stimulating discussions. 
This research has made use of the NASA/IPAC Extragalactic
Database (NED), which is operated by the Jet Propulsion Laboratory, Caltech,
under contract with the National Aeronautics and Space Administration. 
\end{acknowledgements}

\bibliographystyle{aa}
\bibliography{vimos}

\end{document}